# Optical and magnetic resonance imaging approaches for investigating the tumour microenvironment: state-of-the-art review and future trends

**Saumya Prasad,[1,§] Anil Chandra,[1,§] Marta Cavo,[1] Erika Parasido,[2,3] Stanley Fricke,[2,3,4] Yichien Lee,[2] Eliana D'Amone,[1] Giuseppe Gigli,[1,5] Chris Albanese,[2,3,4] Olga Rodriguez,[2,3*] Loretta L. del Mercato[1*]**

[1]Institute of Nanotechnology, National Research Council (CNR-NANOTEC), c/o Campus Ecotekne, via Monteroni, 73100, Lecce, Italy
[2]Department of Oncology, Georgetown University Medical Center, Washington, DC, USA
[3]Center for Translational Imaging, Georgetown University Medical Center, Washington, DC, USA
[4]Department of Radiology, Georgetown University Medical Center, Washington, DC, USA
[5]Department of Mathematics and Physics "Ennio De Giorgi", University of Salento, via Arnesano, 73100, Lecce, Italy

§ equally contributing authors

E-mail: rodriguo@georgetown.edu; loretta.delmercato@nanotec.cnr.it

## Abstract

The tumour microenvironment strongly influences tumorigenesis and metastasis. Two of the most characterized properties of the tumour microenvironment are acidosis and hypoxia, both of which are considered hallmarks of tumours as well as critical factors in response to anticancer treatments. Currently, various imaging approaches exist to measure acidosis and hypoxia in the tumour microenvironment, including magnetic resonance imaging (MRI), positron emission tomography and optical imaging. In this review, we will focus on the latest fluorescent-based methods for optical sensing of cell metabolism and MRI as diagnostic imaging tools applied both in vitro and in vivo. The primary emphasis will be on describing the current and future uses of systems that can measure intra- and extra-cellular pH and oxygen changes at high spatial and temporal resolution. In addition, the suitability of these approaches for mapping tumour heterogeneity, and assessing response or failure to therapeutics will also be covered.

Keywords: tumour-microenvironment, fluorescence microscopy, MRI

## 1. Introduction

Cancer is one of the most lethal diseases worldwide. According to the World Health Organization, cancer is the second leading cause of deaths globally,[1] and according to a report published recently by the American Cancer Society, the overall annual burden is anticipated to rise to 27.5 million new cases being diagnosed with 16.3 million deaths by the year 2040.[2] The complexity of this disease resulting from a multitude of variables remains one of the biggest hurdles in understanding it, and hence, in finding a cure for it.[3,4] Cancer is viewed as a disease consisting of transformed cells that acquire autonomous capacities to undergo limitless proliferation and survival and/or evade the normal programmed cell death through the activation of oncogenes and inactivation of tumour suppressor genes.[5] In the last decade, however, mounting evidence has supported an essential role of the tumour microenvironment (TME) in the development and progression of tumours.[6] The physical, chemical and biological interplay between cancer cells and the TME can determine the course of the disease and therefore TME plays a significant role not only in tumorigenesis, but also in disease prognosis and therapy resistance.[7,8] It is therefore crucial to understand and study in detail the various components of the TME in a spatiotemporal manner. Briefly, the TME is comprised of several cellular and noncellular components.[9] In addition to cancer cell, the cellular components of the TME include cells like fibroblasts, myofibroblasts, neuroendocrine cells, adipose cells, immune and inflammatory cells, along with the vascular and lymphatic vascular networks.[10] The non-cellular component of the TME is comprised by the extracellular matrix (ECM) which includes fibrous proteins such as collagen and fibronectin, other macromolecules such glycoproteins and proteoglycans, and remodelling enzymes such as various metalloproteinases.[11] The ECM not only provides a structural scaffolding, but also promotes and modulates various biochemical and biomechanical processes that affect tissue morphogenesis, differentiation and metastasis. The dynamic interplay between these two components is tissue-specific, highly heterogeneous and defined by tightly regulated conditions that can be divided into two subtypes: physical parameters (pH, partial pressure of $O_2$ and $CO_2$, temperature and the extracellular matrix stiffness)[12] and biochemical parameters (enzymes, adhesion ligands, and secreted factors such as cytokines and growth factors).[13–15] The microenvironment of a tumour is very distinct from that of normal tissues. The metabolism of the TME cells is often deregulated and can be abnormally reliant on aerobic glycolysis (Warburg effect).[16] The TME can also exhibit an aberrant vasculature which can result in poor blood perfusion[17,18] promoting a decrease in pH[19] and decrease in partial oxygen pressure, creating acidotic and hypoxic conditions.[20] Both these factors have been identified as defining hallmarks of tumorigenesis in cancers. For instance, acidosis in the TME has been reported to promote metastasis;[21,22] while the hypoxic response promotes cancer progression by aiding in the adaptation of both cancer and stromal cells to the vicinity microenvironment through hypoxia-inducible factors (HIFs)-dependent signalling mechanisms.[23–25]

The TME provides essential cues for the maintenance of cancer stem cells and for the seeding of cancer cells at metastatic sites. Accordingly, attention to the TME has grown exponentially with many therapeutic strategies now being focused on targeting the TME itself.[26,27] Indeed, targeting hypoxia and acidosis within the TME has emerged as a desirable strategy in the treatment of solid tumours.[28] Therapeutic agents have been designed, which either are selective for molecular targets[29,30] found mainly in hypoxic cells or that can be selectively activated in the hypoxic areas of the tumour.[31] Similarly, with regards to the acidic microenvironment within a tumour, pH-responsive drug-release systems have been developed that are capable of delivering chemotherapeutic agent specifically to acidic tumour microenvironments.[32,33]

Numerous *in vivo* imaging techniques have played a pivotal role in achieving a better and more detailed understanding of the TME, which is a pre-requisite for designing targeted therapeutic agents. For instance, monitoring of extracellular pH has been facilitated by using pH-sensitive fluorescent dyes,[34] pH microelectrodes[35] and inorganic quantum dots.[36] Also, various advanced non-invasive imaging techniques such as Positron Emission Tomography (PET) or Magnetic Resonance Imaging (MRI)-based methods such as Magnetic Resonance Spectroscopy (MRS) and chemical exchange saturation transfer (CEST) MRI have been utilized for this purpose.[37]

In the assessment of hypoxia, several direct and indirect methods have been developed. Direct methods include the polarographic needle electrode[38] which can measure absolute partial pressure of oxygen ($pO_2$); they also include optical imaging using fluorescence/luminescence probes that use different dyes, macromolecules, and nanoparticles[39] and imaging techniques such as phosphorescence imaging,[40] PET,[41] MRI-based Dynamic Contrast-Enhanced and Blood Oxygen Level Dependent[42] and Electron Paramagnetic Resonance.[43] Indirect methods make use of hypoxia exogenous or endogenous markers[44] that are generally semi-quantitative, and include molecular reporters of oxygen such as pimonidaozle and antibody-based immunohistochemistry techniques which can provide important information related to oxygenation of cells.[45,46]

In this work, we will discuss various components of the TME in detail followed by a description of different state-of-the-art techniques employed to measure the most important players of TME, i.e., hypoxia and acidosis. Special emphasis will be given to various imaging techniques employed to measure these parameters. In addition, recent advancements made in this field along with their advantages, limitations and applicability of each technique, will also be discussed in detail with examples.

## 2. The tumour microenvironment

To better understand the imaging strategies that have been developed to study the TME and how these can contribute to the elucidation of the mechanisms of tumour development, progression and behaviour in response to therapeutic interventions, it is necessary to present in detail the main components of TME and how they work.

According to the *Dictionary of cancer terms* developed by the National Cancer Institute (NCI), the TME is the set of healthy cells, blood vessels and molecules that surround and feed a tumour.

Normal cells in the TME, namely, fibroblasts, immune cells and vascular cells, are recruited and reprogrammed by cancer cells in order to support disease progression, through the formation of organ-like structures that are an ideal niche for the promotion of uncontrolled proliferation and local invasion (**Figure 1**). For this process to be successful, a variety of alterations must occur in tumour surrounding tissues that are necessary to assure the survival of the tumour at the expense of normal cells.[40,47,48] In normal conditions, relatively quiescent fibroblasts can become activated in response to disruptions in homeostasis such as in a wound healing response. In cancer, they undergo a similar process, whereby they can deposit a cross-linked collagen matrix that gives origin to fibrosis.[49] Not surprisingly, tumours have been called "wounds that do not heal", a concept introduced by *Dvorak et al.* over 30 years ago[50] and, strengthened since then, by numerous clinical studies that show a clear relationship between chronic inflammation, fibrosis and cancer.[51] Increasing evidence suggests that chronic fibrosis can represent a risk factor for cancer development. For example, idiopathic pulmonary fibrosis (IPF), a chronic fibrotic lung disease, can lead to lung carcinogenesis,[52] while skin fibrosis associated to a particular skin fragility disease known as recessive dystrophic epidermolysis bullosa (RDEB), can be the starting point for highly metastatic skin carcinomas.[53] The same causality can be found in the relation between inflammation and cancer development. For instance, colon carcinoma develops with increased frequency in patients with ulcerative colitis[54] and there is an increased incidence of bladder cancer in patients suffering from chronic Schistosoma infection,[55] two conditions associated with marked inflammation. If fibrosis and inflammation conditions can lead to cancer development, the reverse is also true. Several malignant tumours, such as breast and pancreatic carcinomas, are associated with a very high proportion of fibrotic tissue,[56,57] and all solid tumours present infiltrates of immune cells. In 2008, *Allavena et al.* defined the tumour-macrophage connection as a Yin-Yang relationship, explaining how inflammatory conditions can increase cancer risk (the "extrinsic pathway") and how, on the other hand, genetic events causing cancer (e.g. receptor tyrosine kinase activator) can guide the build-up of an inflammatory microenvironment (the "intrinsic pathway").[58] Fibroblasts associated with cancer, termed as cancer-associated fibroblasts (CAFs) can be distinguished from normal fibroblasts by their unique characteristics.[59,60] From a morphological point of view, CAFs are large spindle-shaped mesenchymal cells with stress fibres and a well-developed fibronexus;[61] they promote tumour progression in several ways, such as by secreting multiple factors such as cytokines and matrix metalloproteinases (MMPs), inducing stemness, promoting epithelial to mesenchymal transition (EMT) and causing epigenetic changes;[62] another fundamental peculiarity of CAFs is their resistance to severe stress such as chemotherapy or radiotherapy, reason why they play a key role in cancer progression.

Even though the link between immune and cancer cells was established many years ago, the significance of this connection has become apparent only recently. Growing literature indicates that immune cells can exert both tumour-suppressive and tumour-promoting effects. For instance, tumour infiltrating leukocytes which can interfere with tumour progression or actively promote it.[63] The immune system, through selection of tumour variants escaping immunologic detection (with reduced immunogenicity), can inadvertently promote tumour growth.[64] This process known as immunoediting consists of three phases: elimination, equilibrium and escape (the "three Es of cancer immunoediting", as defined by *Dunn et al.*).[65]

The first phase of immunoediting, or 'elimination phase', corresponds to the original concept of cancer immunosurveillance, whereby malignant cells are recognized by the immune system and successfully eliminated. Cancer cells that have escaped elimination by the immune system, can become non-immunogenic and be selected for growth in the 'equilibrium phase', where the immune system can control tumour cell growth but is not able to eliminate the transformed cells completely. In this critical phase, the immune system can either eliminate all cancer cells, regressing into elimination phase, or can alternatively select for those cells that showed a less-immunogenic state, promoting the development of the tumour. Cells that are no longer susceptible to immune attack, can result in tumour progression, entering the 'escape phase' of the immunoediting process. Tumours become clinically detectable once cells enter this latter stage.[66]

In this articulated cross-talk between the immune system and cancer cells, both innate and adaptive immune systems are main characters. In the TME, the activation of innate immune cells such as neutrophils, macrophages and dendritic cells (DCs) result in the production and release of lactate, causing a concomitant acidification of the extracellular environment, while activation of adaptive immune cells like T lymphocytes leads to both glycolysis and glutaminolysis, that contribute to production of intra tumour lactate.[67]

The third component of the TME is its microvasculature. Under normal conditions, angiogenesis inhibitors counteract the activity of angiogenesis promoters, keeping the angiogenic switch at an equilibrium. Many diseased states, such as tumours, are characterized by abnormally low $pO_2$, low pH, hypoglycaemia, and increased immune/inflammatory stimuli and that cause a dramatic increase in endothelial proliferation; thus, the "angiogenic switch" is turned on.[68] The formation of new blood vessels from pre-existing capillaries begins with the detachment of pericytes and dilation of blood vessels, allowing endothelial cells to migrate into the perivascular space towards angiogenic stimuli; guided by the pericytes, endothelial cells proliferate and migrate, and assemble to create a lumen, which is followed by basement-membrane formation and pericyte attachment. In the end, various blood-vessel sprouts fuse with one another to build a new circulatory system.[69]

It is now accepted that cancer-associated cells (i.e. CAFs, infiltrating immune cells and angiogenic vascular cells) play a central role in enabling the various hallmark capabilities of cancer. In 2012, *Hanahan and Coussens* defined this complex population of cells as "accessories to the crime"; the authors identified eight fundamental characteristics of cancer, seven of which demonstrably involved the contribution by TME stromal cells. Thus, CAFs, infiltrating immune cells and angiogenic vascular cells were found to be involved in (i) sustaining proliferative signalling, (ii) resisting cell death, (iii) activating invasion and metastasis and (iv) avoiding immune destruction; CAFs and infiltrating immune cells also contributed in (v) inducing angiogenesis and (vi) evading growth suppressors; CAFs alone were proven to contribute in (vii) deregulating cellular energetics, while no contributions from stroma cells were found in (viii) enabling replicative immortality, the last cancer hallmark.[70]

Tumour angiogenesis differs significantly from physiological angiogenesis: it exhibits an aberrant vascular structure, altered endothelial-cell–pericyte interactions, abnormal blood flow, increased permeability and delayed maturation. Consequently, it is intuitive that tumour blood vessels are different from their normal counterparts: they are irregularly shaped, tortuous and poorly organized, and the resultant vascular network is often leaky and haemorrhagic;[71,72] moreover, tumour vessels have also been reported to have cancer cells integrated into the vessel wall.[72,73]

Stroma cells and blood vessels synthesize and release growth factors, chemokines and adhesion molecules (**Figure 1**), that affect the homeostasis and composition of the TME, making it heterogeneous between different tumours.[74,75] Extracellular proteinases, such as matrix metalloproteinases (MMPs), mediate many of the changes of the TME, playing a crucial role in various physiological processes such as organ development[76] and in tumour behaviour through their remodelling capabilities and as key players in the molecular communication between tumour and stroma. Thus, MMP inhibitors have been tested, and mostly failed, as anticancer drugs; the reason being their capacity to either suppress or promote tumorigenesis,[77] depending on the cell context.

All the above-mentioned factors contribute to modifying the physicochemical and metabolic conditions within the TME, making it considerably different from the microenvironment found in healthy tissues. Basically, tumours are characterized by lack of oxygen (hypoxia), lower pH values (acidic conditions) and low glucose levels (hypoglycaemia).

Areas with very low (down to zero) oxygen partial pressures can exist in solid tumours, mainly because oxygen delivery to cells is affected by a deteriorating diffusion milieu, severe structural abnormalities of tumour capillaries and disturbed microcirculation.[78] These hypoxic microregions are heterogeneously distributed within the tumour mass and may be located adjacent to regions with normal $O_2$ partial pressures (**Figure 1**).[79]

The hypoxic microenvironment in solid tumours has profound effects on tumour propagation and evolution: on one side, hypoxia can induce changes in the proteome of neoplastic and stroma cells, resulting in cellular quiescence and apoptosis; on the other, hypoxia-induced proteome changes can also promote tumour propagation by enabling cancer cells to overcome the nutritive deprivation by promoting the generation of new blood vessels, that allow cells to escape and spread into the body.[80]

Hypoxia forces a glucose metabolism detour through the glycolytic pathway instead of respiration, resulting in the production of lactic acid that acidifies the intra-tumoral environment.[81–83] As a consequence, the extracellular pH ($pH_e$) of tumours tends to be below 7.0, while the intracellular pH ($pH_i$) is maintained at a neutral range (>7.0) similar to that of healthy tissues (**Figure 1**). This gradient between $pH_e$ and $pH_i$ renders the cells resistant to weakly basic drugs by hindering their cellular uptake, whereas it increases the uptake of weakly acidic drugs.[84]

Apoptosis or necrosis of tumour cells can occur in response to hypoxia. Necrosis, in particular, causes an egress of intracellular ions, such as $K^+$, into the extracellular compartment leading to an ionic imbalance that can contribute to T cell dysfunction.[85]

These metabolic changes have been associated with tumour progression and poor clinical prognosis. The mechanisms behind these effects involve the induction of the hypoxia-inducible factor (HIF) family of transcription factors, that consists of three members, namely, HIF-1,-2, and -3, which act to regulate cellular processes including glucose metabolism, cell proliferation and tissue remodelling in response to hypoxic conditions.[86]

With the complexity of the TME in mind, in order to effectively translate this knowledge into effective therapeutic strategies, it is also necessary to consider the physical and mechanical structure of tumours, that can deeply affect the interstitial fluid pressure, which can in turn hinder the penetration of drugs through tissues[87] and can negatively affect the sensitivity to both radiation and chemotherapy.[88] For this reason, three-dimensional (3D) systems are necessary.[89–93]

Indeed, while bi-dimensional (2D) monolayers have been the cornerstone of preclinical cancer research, cells grown in monolayers do not accurately reflect the biological complexity of tumours. Solid tumours grow in a 3D conformation that expose cells to very specific conditions, such as hypoxia and a heterogeneous distribution of nutrient levels, which can affect cell fate.[94] In conventional monolayer cell cultures, most of these environmental cues are missing and consequently, 2D systems often yield unsuccessful and contradictory results.[95] Conversely, 3D cultures more accurately recapitulate *in vivo* cell-cell interactions, thus the importance of 3D tumour cell cultures in the study of cancer pathogenesis and evolution, as well as in the evaluation of drug efficacy, is being increasingly recognized by the scientific community.[96–102]

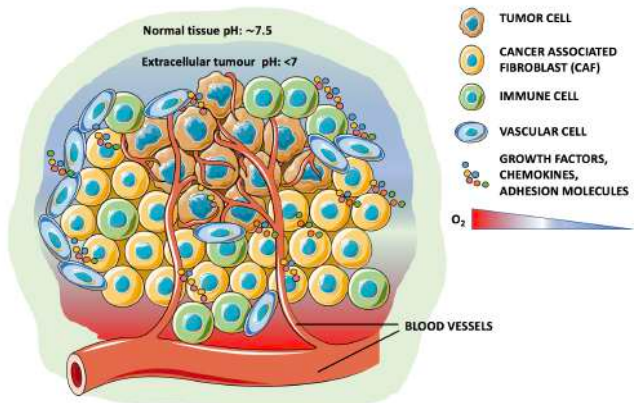

**Figure 1 – Schematic representation of the tumour microenvironment (TME).** In addition to cancer cells, the TME is composed of cancer associated fibroblasts, immune cells and vascular cell, growth factors, chemokines, adhesion molecules and blood vessels. All these components contribute in modifying the physicochemical and metabolic conditions with respect to those of healthy tissues; in particular, the lack of oxygen (hypoxia) and lower pH values (acidosis) are characteristic of the TME.

## 3. Optical imaging of tumour cells and TME

Light-based detection and sensing systems have been used for tumour visualization and sensing of its different microenvironment components such as estimation of acidity and hypoxia, ion concentrations, lactate and expression of certain proteins of interest.[37,39,44,103] A low pH[22,34,104,105] and dissolved oxygen concentration (hypoxia)[23,25,106,107] are the two most common features of the TME and play a major role in imparting resistance of cancer cells to chemotherapeutic agents. Various techniques have been utilized for monitoring these parameters in *in vitro* and *in vivo*, as well as for assessing tumour heterogeneity and drug response. The most widely used imaging techniques in the clinical setting include MRI, Computed Tomography (CT) scan, PET, however they present, limitations in terms of either resolution and non-specificity, as well as invasiveness and cytotoxicity, due to the use of radioisotopes or harmful high energy radiation.[108] Fluorescence-based imaging has been considered as an alternative because of its simplicity, high resolution and non-damaging effect on cells and the body. In this section, we will mainly discuss the fluorescence-based sensors which have been routinely applied for monitoring acidosis and hypoxia in tumours, along with their advantages and limitations.

Optical imaging is an inexpensive bioimaging tool that provides an excellent in-plane resolution (50 -200 nm)[109,110] and sensitivity with a high temporal resolution (200-16000 frames per second)[111], and provides great versatility which allows its tailoring according to the experimental requirements. Due to its high sensitivity, it can detect very low concentrations of target analytes in the range of nanomolar concentrations.[37] Additionally, its temporal resolution nowadays is very high and improving continuously.[108,112] One of the most useful techniques for *in vitro* analysis is fluorescence confocal microscopy. It is very suitable for acquiring high-resolution images of cellular architecture, including both intra- and extracellular locations.[113] Spatiotemporal imaging flexibility and choice of spectral wavelength are two tremendous advantages in fluorescence imaging systems, which are often used for imaging tumour spheroids with a very high resolution.[114] However, one of the drawbacks of fluorescence microscopy is its limited depth of penetration in thick biological samples, where phenomena like absorption, reflection and refraction can significantly block the excitation and emitted light.[115] Thus, light microscopy has practical limitations in the visualization of thick tissue samples and in the imaging of live animals that hinder the measurement of *in vivo* parameters such as its tumour pHe. Furthermore, weaker optical signals originating from deeper locations of surface-accessible tumours can be influenced due to depth and thus result in the appearance of a modified characteristics of the tumour surface, which may be different from the actual scenario. Due to these reasons, the measurement of pHe with light-based bioimaging has not been practiced in routine clinical applications.[116] Sensitive probes in the near infrared region of the spectrum have been used to tackle these limitations, as these probes are less affected due to lower scattering and absorption that allows them to penetrate deeper into the tissue.[117] Considering the bottlenecks and practical limitations of fluorescence microscopy, newer and improved fluorescent probes and microscopy techniques are continuously being devised. Thus, a continuous search for more innovative use of the available probes can fill the gap and result in clinically viable platforms. For a comparative understanding about these sensors and bioimaging agents, a few important properties, advantages and shortcomings are represented in tabulated form (Table 1).

### 3.1 Imaging acidosis within the TME

As discussed in the earlier sections, the pHe of a tumour (6.2-6.9) is lower than that of healthy tissues (7.4). This property of tumours can be used to gain a deeper understanding of cancer and to develop chemotherapeutic drugs that can target either the acidity of the TME per se or, areas of acidosis where cancer cells reside.[118] Towards this end, various imaging techniques have been developed and used to selectively sense the low pH around a tumour. In this section, we will discuss the various optical probes that have been employed to monitor low pH within the TME. Ideally speaking, a pH-sensitive imaging probe should (i) have negligible fluorescence in the extracellular environment (blood or interstitial fluid) surrounding healthy tissues, i.e. at pH 7.4; (ii) exhibit high fluorescence in acidic conditions (pH < 6.3); (iii) emit fluorescence in the NIR range for a better signal-to-noise ratio and exhibiting less autofluorescence for *in vivo* applications; (iv) have tuneable pKa values for multiple applications.[119,120]

Fluorescence-based imaging sensors used for assessing the cellular microenvironment depend on the use of fluorescent dye molecules, quantum dots and fluorescent nano- and microparticles.[121–123] The fluorescence in the nano- or microparticles which are not intrinsically fluorescent is generally produced upon triggering excitation of small fluorescent molecules which are incorporated in the particles and are which are sensitive towards a target parameter.[124–126] On the other hand semiconducting quantum dots and fluorescent carbon based quantum dots are intrinsically fluorescent due to quantum confinement effect and do not need labelling with fluorescent dye molecules. However, Tto increase the sensitivity and to nullify the effect of concentration on the fluorescence readout, a second reference fluorescent signal is typically required, which is insensitive to the parameter of interest. The combined use of a sensitive and an insensitive signal is known as ratiometric sensing, which is very useful and is applied frequently in *in vitro* cell analyses. Broadly, pH-responsive probes can be classified into the following categories:

### a) Fluorescent pH-responsive small molecules: Several small-molecule probes which become fluorescent only under acidic pH conditions, have been widely used to monitor both the extracellular and intracellular pH of tumours.[127,128] These probes are constructed such that they exhibit a high specificity and sensitivity for the highly acidic TME due to protonation. Additionally, they have a high target-to-

background signal ratio with minimum background contamination coming from the non-target/healthy tissues. The fluorophore in these probes could be conjugated with a peptide[129] or antibody[130] to enhance selectivity for the tumour tissue. The method of analysis is usually ratiometric as it offers pH monitoring independent of the dye's concentration. The pH could be assessed either by monitoring the ratio of two signals emitted at different wavelengths[131,132] or by comparing fluorescence lifetime of a fluorescent molecule/sensor under different conditions.[133] A well-known example of ratiometric imaging is seminaphtharhodafluor (SNARF)[132] which was used by *Anderson et al.*[134] to conjugate this ratiometric pH indicator to the Lys residues of a pH low insertion peptide (pHLIP) for targeted measurement of cell surface pH both *in vivo* and *ex vivo*. They studied the effect of this probe on both highly metastatic and non-metastatic tumour spheroids, where they found that the average surface pH values of highly metastatic tumours were lower than the non-metastatic tumours. Notably, the method was sensitive enough to detect 0.2–0.3 pH unit changes *in vivo* in tumours induced by glucose injection. Recently, *Xu* and co-workers[135] have described a two-photon high-resolution ratiometric probe for imaging of acidic pH in living cells and tissues in the early detection of tumours in a mouse model. They developed two carbazole−oxazolidine-derived fluorophores (PSIOH and PSIBOH) in which pH sensing was based on the opening of the oxazolidine ring upon protonation (**Figures 2a** and **2b**). PSIOH exhibited large spectral shifts of ~169 nm under acidic conditions with a pKa of 6.6, making it ideal for monitoring and imaging the intracellular and extracellular pH changes using two-photon microscopy (**Figure 2c**). The probe exhibited a high signal-to-noise ratio with an outstanding photostability and reversibility, and low cytotoxicity. Using this probe, the authors were able to detect pH changes (caused by exogenous stimulation) in HeLa cells, as well as detect and

| Table 1. Important advantages and disadvantages of different fluorescent sensors | | |
|---|---|---|
| | **Advantages** | **Disadvantages** |
| **Small molecule (Few angstrom)** | • Convenient to use<br>• Minimum perturbation to the interrogated system<br>• A large number of molecular probes are available<br>• Lifetime: 0.1 – 20 ns, up to 90 ns (pyrenes) | • Poor photostability [136]<br>• Low stability in biological environment [137]<br>• Poor hydrophilicity [138]<br>• Rapid elimination from body |
| **Semiconducting Quantum dots (2-10 nm)** | • Intrinsically fluorescent with high quantum yield (~95%)<br>• Resistance towards photobleaching<br>• Large stokes shift and size dependent fluorescence properties<br>• Broad absorption range and long fluorescence lifetimes (average – 10-30 ns, up to 500 ns) | • Highly cytotoxic degradation byproduct due to heavy metal based composition [139]<br>• Show photoblinking<br>• Known to produce free radicals during excitation [140]<br>• Size dependent accumulation in organs [141,142] |
| **Polymer dots (<100 nm)** | • Tailorable fluorescence properties<br>• Easy to bioconjugate<br>• Multiple sensing possible by incorporating different sensing molecules | • Size dependent accumulation in organs and tissues [143]<br>• Broad emission spectra [144] |
| **Organic quantum dots (carbon dots, <10 nm)** | • Lower cytotoxicity compared to semiconducting QDs<br>• Photobleaching resistant<br>• Tunable fluorescence properties<br>• Fluorescence lifetime: ~5 to 20 nanosecond | • Size dependent accumulation in organs<br>• Known to generate reactive oxygen species [145–147] |
| **Gold nanoclusters (< 3 nm)** | • Highly sensitive [148–151]<br>• Less cytotoxic<br>• Excellent photostability and good water dispersibility<br>• Size tunable emission maxima<br>• Tunable fluorescence lifetimes (Hundreds of nanoseconds to microseconds) | • Unidirectional sensing (Irreversible)<br>• Relatively low quantum yields [152]<br>• Sensitive to environment |
| **Gold nanoparticles (~10-100 nm)** | • Highly Sensitive FRET based sensors can be developed<br>• Low cytotoxicity | • Intrinsically Non-Fluorescent (Dye has to be linked) [153]<br>• Prone to aggregation<br>• Size dependent accumulation in organs [154,155] |
| **Fluorescent proteins (covalently labeled)** | • Brightness comparable to organic dyes<br>• More stable and very less photoblinking<br>• Lifetimes: 0.1 – 4 ns | • Low specificity compared to unlabeled proteins<br>• Decreased intracellular concentration after multiple passage causing decreased signal to noise ratio. [156] |
| **bio-reductive probes** | • Highly specific hypoxia and reductase enzyme sensing<br>• Turn on sensor (high sensitivity) | • Reversible bio-reductive probes are rare [157] |
| **Metal-ligand complexes and metalloporphyrins** | • Long lifetimes: hundreds of ns to tens of μs<br>• Very high signal to noise ratio possible due to longer lifetimes<br>• Photostable<br>• Large stokes shift | • Cytotoxicity due to nonspecific binding to membrane and proteins [158]<br>• Nonspecific sequestration inside the cell |
| **Genetically encoded proteins** | • High selectivity and sensitivity,<br>• amenability to subcellular targeting, and applicability to tissue-specific or whole-body imaging in animals.<br>• Lifetime: < 2.25 ns | • High expression of sensor proteins may alter normal function of cells [159]<br>• Variation in expression among different cells<br>• Poor photophysical properties<br>• Low quantum yield and limited color |
| **Endogenous fluorophores** | • No need of external sensor<br>• Lifetime: 0.1 – 7 ns<br>• No processing of specimen<br>• Useful for optical metabolic imaging | • Complex data analysis required [160]<br>• Low penetration depths due to use of UV for excitation of NADH and FAD molecules |

distinguish cancer and tumours in the liver of mice, thus, making it appropriate for both *in vitro* and *in vivo* applications. Another classic example in this category includes an activatable probe developed by conjugating the 2,6--dicarboxyethyl-1,3,5,7-tetramethyl-boron-dipyrromethene (BODIPY) fluorophore to a cancer-targeting antibody (trastuzumab). For instance, *Urano et al.*[120] developed a series of pH-responsive probes with sensitivity to pH in the range of 3.8-6.0. These probes were based on photon-induced electron transfer (PeT), wherein the probes were non-fluorescent in the non-protonated form, while they were highly fluorescent in acidic conditions with emission at ~500 nm **(Figure 2d)**. In addition, the activation of these probes was reversible, which produced signals only in viable cancer cells.

For *in vivo* imaging, the NIR range (650–900 nm) is usually preferred over photons that emit in the visible range, because they can propagate deeply through tissue (approximately 1 cm) [37] while exhibiting low tissue absorption and autofluorescence.[161] Thus, this optical *in vivo* window has been extensively exploited for the development of several probes. An NIR pH-activatable probe has been developed by *Lee et al.*[162] by conjugating a pH-sensitive cyanine dye to a cyclic arginine-glycine-aspartic acid (cRGD) peptide targeting α,β₃ integrin (a protein that is overexpressed in endothelial cells during tumour angiogenesis) for *in vivo* and *ex vivo* imaging of orthotopic 4T1/luc tumours in a mouse model of breast cancer **(Figures 2e** and **2f)**. This probe (pKa of 4.7) had a low background signal with negligible fluorescence above pH 6.0, thus making it an ideal probe for imaging pH in endosomes and lysosomes.

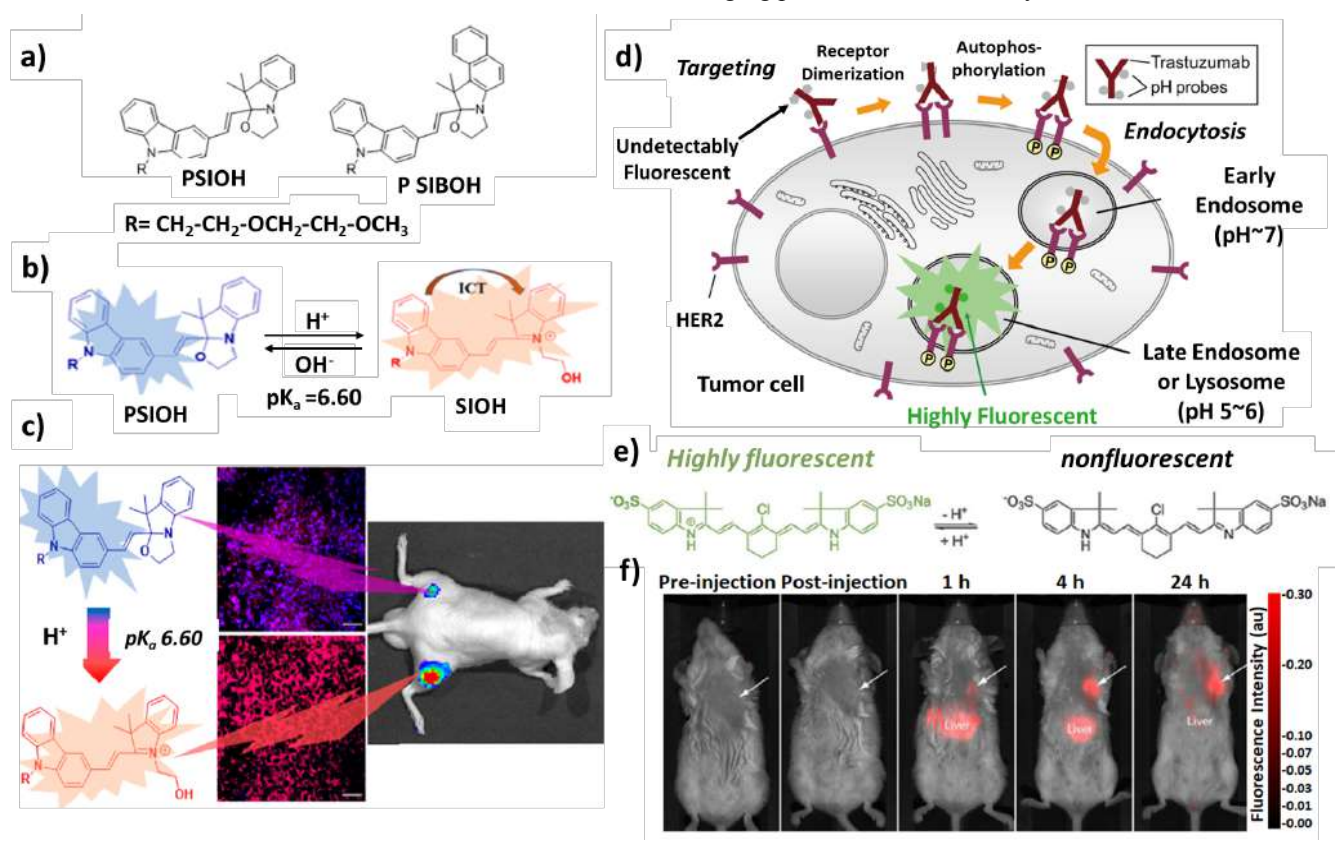

**Figure 2 – Small molecule-based-pH-sensing probes for optical imaging. (a)** Structures of PSIOH and PSIBOH; **(b)** pH sensing mechanism of PSIOH; **(c)** Imaging of tumour-bearing mice injected with PSIOH;[135] **(d)** Schematic representation of tumour imaging mechanism by BODIPY- trastuzumab conjugate;[120] **(e)** Structure of pH-sensitive cyanine dye in acidic and basic conditions; **(f)** Structure of cRGD-linked cyanine dye and its application in imaging of orthotopic breast cancer-bearing mice at different time points (arrows indicate the tumour region).[103,162]

Depending on the availability of different targeting groups, probes based on small molecules could be widely utilized to image the TME specifically. However, there are only a few dyes which 'turn-on' upon protonation to produce fluorescent signals; therefore, other small molecule-based fluorescent pH probes should be developed for the advancement of tumour bioimaging and microenvironment monitoring.[103] In addition to the NIR-I based fluorescent probes, which provides the first biological window, NIR-II (1000- 1700 nm) provides additional leverage and know as second biological window. The penetration depth of imaging could be in centimeters and high resolution imaging could be achieved at millimeter depths. It has been observed that compared to NIR-I, autofluorescence from tissue of endogenous fluorophores is significantly reduced with NIR-II. Therefore the signal to noise ratio is significantly reduced with NIR-II based probes.[163,164]

A recent example of NIR-II based sensing is work by Wang et al, where they develop a series of antiquenching pentamethine cyanine fluorophores that exhibit stable absorption/emission beyond 1000 nm (peak absorption/emission up to 1015/1065 nm in aqueous solution) with up to ~ 44 fold enhanced brightness and superior photostability in aqueous solution. These fluorophores showed deep optical penetration (~8 mm) with high contrast and stable lymphatic imaging. In fact they were observed to be better than clinical-approved ICG, hence are suitable for clinical translation. In addition, due to its ratiometric pH sensitivity imaging and quantification of gastric pH was also accomplished with reliable accuracy and upto depths of 4 mm. [165]

Recently Tian et al showed multiplexed NIR-II based probes for lymph node invaded cancer detection and imaging guided surgery. The two probes were employed to separately visualize metastatic tumor and the tumor metastatic proximal LNs resection. They used nonoverlapping NIR-II probes that showed significantly suppressed photon scattering and zero autofluorescence. They used tumour seeking donor-acceptor-donor (D-A-D) dye, IR-FD in NIR-IIa window (1100-1300 nm) The other fluorophore was PbS/CdS core/shell quantum dots (QDs) with dense polymer coating. They were used to visualize cancer-invaded sentinel LNs in the NIR-IIb (>1500 nm). The probe exhibited superior brightness and photostability even after continuous irradiation for 5 hours and works with picomolar dose for sentinel LNs detection. The authors claimed that the probes can be used even under bright light that is beneficial under clinical setup.[166]

In another work Wang et al showed applicability of NIR-II based nanoprobes in-vivo assembly to improve image-guided surgery for metastatic ovarian cancer. The probes based on downconversion nanoparticles (DCNPs) modified with DNA and targeting peptides showed improved image-guided surgery against metastatic ovarian cancer. Compared to ICG the imaging quality was superior with good photostability and tissue penetration. The authors showed that metastases with ≤1 mm can be completely excised under NIR-II bioimaging guidance.[167]

*b) Fluorescent pH-responsive nanoprobes:* Nanoprobes have various advantages when compared to small molecular probes such as the capacity to achieve increased accumulation in the tumour due to enhanced permeability and retention (EPR) effect, high signal-to-noise ratio, and tuneable circulation lifetime with predictable pharmacokinetics. Additionally, enhanced specificity for the target tissue can be attained by labelling numerous imaging molecules on a single multivalent nanoparticle.[168] Fluorescent nanoprobes can be designed to have an enhanced signal to noise ratio, where there are different mechanisms to achieve it. One of the simplest techniques is enhancing brightness by using intrinsically bright sensors based on semiconducting quantum dots[169] or by accumulation of large number of fluorophores (fluorescent small molecules) into the nanoprobe.[170] The signal produced from one small molecule is very less and thus in a solution the overall signal density at any one point is very less. In the nanosensors consisting of confined multiple small sensing molecules, the overall brightness is significantly higher and thus random fluctuations in signal output from individual molecules is nullified that results in heightened signal to noise ratio. In biological specimens such as cells and tissue mass the scattering effect can cause serious reduction in signal.[171] Utilization of near IR excitation and/or emission wavelengths is another effective approach in enhancing signal. Use of a higher wavelength reduces the scattering when the excitation signal is traversing through the sample, in a similar fashion the emitted signal scattering is also less for higher wavelength light.[172] Two photon microscopy, where a fluorophore can be excited using higher wavelength photon, is another technique for getting higher signal to noise ratio. It is beneficial as intrinsic fluorescence from the specimen is strongly diminished because of their poor excitation. The fluorophore or the sensor on the other hand can be excited in a small volume, where it can simultaneously absorb two photons that together provides enough energy for excitation.[173] Here the probability of this event is very unlikely outside the excitation cone that further reduces the noise from the sample and enhances the contrast in two photon microscopy.

The pH-activatable nanoprobes are based on mechanisms like Förster Resonance Energy Transfer (FRET) and the Self-Aggregation Associated Energy Transfer (SAET). The nanoprobes can be based either on inorganic material (e.g. quantum dots)[174] or biomaterial such as dextran[161] or synthetic polymers (e.g. dendrimer-based NIR fluorescent nanoprobe).[175] Several applications when quantum dots have been used as FRET donors have been described by various groups.[176,177] In addition, pH nanoprobes based on SAET have also been developed.[178] An exhaustive review on pH-responsive fluorescence nanoprobes for tumour visualization by *Wang* and *Li* summarizes various other examples of nanoprobes used in the past.[168] Here, some of the recent advancements in the use of nanoprobes for monitoring the pH of TME will be discussed in detail.

*Ding et al.*[179] described NIR glutathione and mercaptopropionic acid co-modified Ag$_2$S nanodots (GM-Ag$_2$S NDs) which act as acceptors of energy, and up-conversion nanoparticles (UCNPs) as the energy donors for sensing and imaging pH *in vivo* **(Figure 3a)**. The fluorescence up-conversion involves the use of two or more higher wavelength photons to cause excitation, that results in fluorescence emission at a lower wavelength. Therefore, up-conversion is very suitable for deep tissue imaging and sensing applications.[180] In this case, the nanodot/nanoparticle FRET pair works based on ratiometric fluorescence up-conversion, where luminescence resonance energy transfer (LRET) takes place. This LRET-based nanoplatform showed potential for the construction of up-conversion luminescent nanoprobes for biological applications. In another recent report, *Yu et al*.[181] described the application of multifunctional gold nanoparticles as smart nano vehicles for intracellular pH mapping and *in vivo* magnetic resonance/fluorescence imaging (MRI/FI). They synthesized two types of pH-sensitive MRI/FI dual-modal nano vehicles, where one type (Au@Gd and Au@Gd&RGD) was modified with a rhodamine derivative, which showed sensitivity towards acidic conditions and could be used to measure pH variation of lysosomes in living cells. The second type of nano vehicle (D-Au@Gd and D-Au@Gd&RGD) consisted of two pH-responsive fluorophores, derived from rhodamine and fluorescein, which were able to quantitatively assess the pH values of living cells **(Figure 3b)**. A successful *in vivo* MR and fluorescence imaging of D-Au@Gd&RGD in tumour-bearing mice proved the ability of the nano vehicles to penetrate the blood–brain barrier (BBB), essential for applications related to central nervous system (CNS). Thus, this imaging agent could be used as a diagnostic tool for determining location and size of tumours and to monitor their response to treatment *in vivo*. In a different approach, *Reichel et al* prepared a nanoprobe for cancer imaging by combining two FDA-approved agents: feraheme (FH) and indocyanine green (ICG).[182] The resulting FH-entrapped ICG nanoprobe [FH(ICG)] exhibited quenched fluorescence emission, which was re-activated in the presence of acidic TME conditions (pH 6.8) in cancer cells. *In vivo* studies in a prostate cancer mouse model showed that FH(ICG) administration enhanced the long-term fluorescence signals in tumours compared to ICG alone **(Figures 3c** and **3d)**. Advantages of these probes include their activatable nature and high signal-to-noise ratio with long-term fluorescence signals that enabled their fluorescence-guided detection.

Therefore, this nanoprobe offers tremendous potential as a clinically translatable image-guided cancer therapy system that could be used in a clinical setting.

Linking of small fluorescent-based sensors with nanoparticles can improve their biocompatibility, delivery efficiency and, sensitivity by permitting the adjustment of their emissions within a useful range. Therefore, the development of pH nanoprobes greatly diversifies the applications of pH-sensitive small molecule dyes in the optical monitoring of the TME. Specificity can be endowed to these nanosensors by attaching targeting groups onto their surface.[183,184] These targetting groups could be protein, antibodies or nucleic acids linked to the nanosensor by covalent or physical interaction. Antibodies being the most commonly used for imparting biospecificity can be used in different forms like intact antibodies, Fab, F(ab)₂, Diabodies and Minibodies. The choice depends on the desired application, for example Fab fragments can be used for rapid tumor targetting and penetration due to its small size but has low avidity. Intact antibody on the other hand has high specificity towards antigens but show long circulation times, high background and are expensive. Similarly other forms of antibodies have their own advantages and shortcomings.[185]

The majority of these tumor-targeting entities are used for tagging cell membrane-bound proteins. Different categories of targetable membrane-bound proteins include anchoring proteins, receptors, enzymes, and transporter proteins. The functions and biological properties of these proteins determine their position and distribution on the cell membrane, that can cause variability in their availability. Out of the ~7,000 known transmembrane proteins about 150 are overexpressed on tumor cells or tumor-associated vessels, thus they can be used for therapeutic targeting or imaging.[186]

Although the pH nanosensors are very tailorable ~~However~~, they ~~Specificity can be endowed to these nanosensors by attaching targeting groups onto their surface. However, they~~ can be somewhat limited in their applications due to their specific and narrow pH range and, their big size which could cause a high uptake in the liver and kidney, when performing *in vivo* imaging. Therefore, the development of more biocompatible and bioresorbable nanocarriers may help solve these problems for *in vivo* applications.[103]

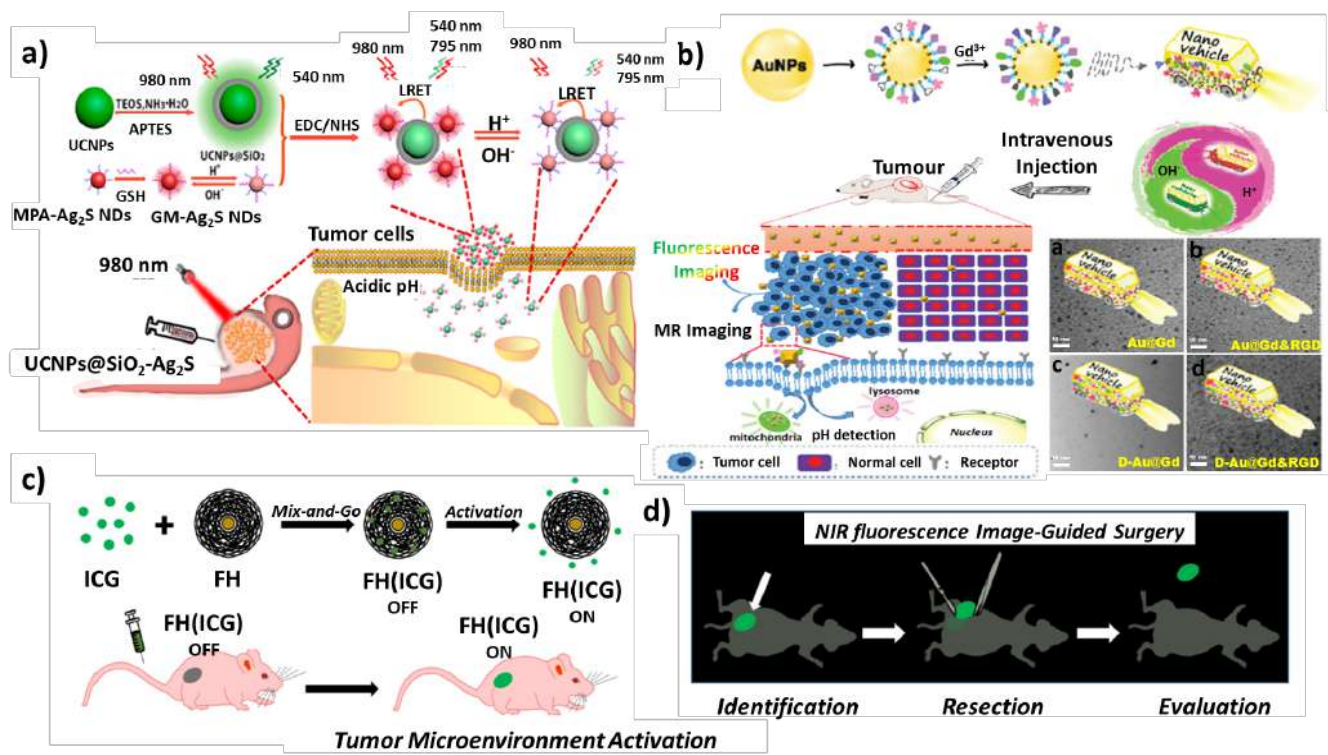

**Figure 3 – pH-responsive nanoprobes. (a)** Schematic of LRET-based nanoplatform as a pH-responsive ratiometric nanoprobe;[179] **(b)** Schematic representation of the preparation of nano vehicles with TEM images and their application for *in vivo* MR/fluorescence imaging;[181] **(c)** Schematic illustration of the mechanism of activation of FH(ICG) for cancer imaging; **(d)** Graphic representation of image-guided NIR surgery with FH(ICG) in a mouse model.[182]

*c) pH-responsive fluorescent proteins:* Biosensor fluorescent proteins are exciting probes that can aid in the monitoring of biological features and present advantages such as having minimized interference from the endogenous chromophores in cell cultures or animal models. In this context, proteins such as green fluorescent protein (GFP) and its variants have found wide applications as indicators of intracellular pH[187,188] due to its robust fluorescence properties, site-specific labelling, pH sensitivity, and mutations for multiple-site labelling.[189] Also, since they are genetically encoded, they can be used for the specific measurement of subcellular pH.[168]

We will discuss some of the recent advancements in the imaging of tumour intracellular pH using fluorescent proteins. Ratiometric intracellular pH imaging was successfully performed in cancer cells (HeLa Kyoto) expressing SypHer2, a fluorescent indicator, both *in vitro* (2D cultures and tumour spheroids) and *in vivo*.[190,191] *Burgstaller et al.* have recently reported a fluorescent protein-based pH reporter (pH-Lemon) for acidic cellular compartments like endosomes and lysosomes.[192] They fused mTurquoise2 (a pH-stable cyan fluorescent protein variant) with a highly pH-sensitive enhanced yellow fluorescent protein (EYFP), yielding a freely reversible and ratiometric biosensor named pH-Lemon. This ratiometric probe is based on the FRET mechanism and under acidic conditions, the fluorescence of EYFP decreases while that of mTurquoise2 increases **(Figures 4a** and **4b)**. Additionally, the authors fused this sensor with specific markers, which allowed its targeting to specific organelles of HeLa cells, making it very suitable for the study of local pH dynamics of subcellular microstructures in living cells **(Figure 4c)**. In another report, *Tanaka et al.* described a novel method to visually determine the intracellular tumour pH in a xenograft model *in vivo* by ratiometric imaging of fluorescent proteins.[193] The yellow fluorescent protein (YFP, pH-sensitive) was used as the pH indicator, whereas the red fluorescent protein (RFP, pH-insensitive) was used as the reference **(Figure 4d)**. Tumours in nude mice were xenografted with HeLa cells expressing the RFP-YFP probe, showed a heterogeneous distribution of different intracellular pH in the tumour tissue **(Figure 4e)**. Additionally, using angiography performed with an NIR fluorescent probe, allowed them to correlate the changes in intracellular pH with angiogenesis. Other advantages, according to the authors, included eliminating the influence of optical factors from tissue, as well as only requiring filter-based fluorescent imagers commonly used in small animal studies.

### 3.2 Imaging hypoxia within the TME

Oxygen is essential for energy metabolism and for cellular bioenergetics. When the oxygen demand of rapidly growing tumours exceeds the amounts that can be provided by their surrounding vasculature, a drop of normal oxygen levels from 2–9% to hypoxic levels of less than 2% results in hypoxic regions. The clinical significance of hypoxia in cancer therapy has been reported in several extensive reviews.[20,79,194] Hypoxia is yet another hallmark of solid tumours and is considered a critical TME parameter. The progression of cancer, its responsiveness to treatment and development of resistance, are known to be strongly affected by hypoxia in addition to other microenvironmental factors like pH.[195–198]

Due to the inevitable and widespread hypoxia present in all kinds of cancerous tissues, its high resolution and accurate imaging is of great importance in the diagnosis and evaluation of therapeutics on tumours. Therefore, the development of non-invasive, sensitive and specific molecular imaging methods for the evaluation of hypoxia is a central focus of current cancer research.

In the past, several fluorescent imaging probes have been developed that enable non-invasive imaging of the hypoxic microenvironment.[199] An ideal probe for sensing hypoxia should have several characteristics, i) It should be activatable only in hypoxic conditions and not in normoxic conditions, ii) Should show reversibility, iii) Should be independent of co-variable parameters such as pH and blood flow, iv) Mechanism of cellular retention should be well defined and cell type independent, v) Balanced lipophilic/hydrophilic property uniform tissue distribution, and to avoid membrane sequestration, and have faster clearance from systemic circulation and normoxic tissue, vi) High stability against non-hypoxia specific metabolism in vivo, vii) Amenable dosimetry profile viii) Should work on multiple tumour types, ix) Should be easy to synthesise and readily available[200]

Several activatable (on/off) probes based on mechanisms such as fluorescence, phosphorescence and FRET have been developed.[201–203] The hypoxia sensing mechanisms rely either on redox-triggered changes in fluorescent intensity of exogenous probes or on the oxygen-induced quenching of phosphorescence of exogenous probes.[199] These probes could be nanoparticles (NPs), metal-ligand complexes or fluorescent proteins. The readers are directed to an exhaustive review by Wang and Wolfbeis for more detailed information on optical methods for sensing and imaging oxygen.[204] In this section, we will discuss the recent developments in optical imaging probes used for targeting hypoxia in the TME.

*a) Redox sensitive bio-reductive probes:* Low oxygen concentration in the TME may cause enzymatic bioreduction and consequent selective entrapment of small amphiphilic molecules, such as 2-Nitroimidazole and indolequinone in hypoxic cells, a feature that has been exploited to construct fluorescent probes for imaging of intratumoural hypoxia.[157,205] *Okuda et al.* developed a novel NIR fluorescent probe (GPU-167) which is a conjugate of 2-Nitroimidazole and the NIR dye Tricarbocyanine for *in vivo* imaging of tumour hypoxia.[203] They were able to demonstrate the specificity of this probe for tumours and thus its potential as an *in vivo* optical imaging probe for tumour hypoxia (**Figures 5a** and **5b**). A similar approach was used by *Biswal et al.,* where they used a 2-Nitroimidazole bis-carboxylic acid indocyanine dye conjugate for tumour-targeted fluorescence imaging of hypoxia.[206] In another report involving use of 2-Nitroimidazole , Liu and colleagues used novel fluorescent markers containing naphthalimides and two 2-Nitroimidazole chains for imaging of hypoxic cells.[207]

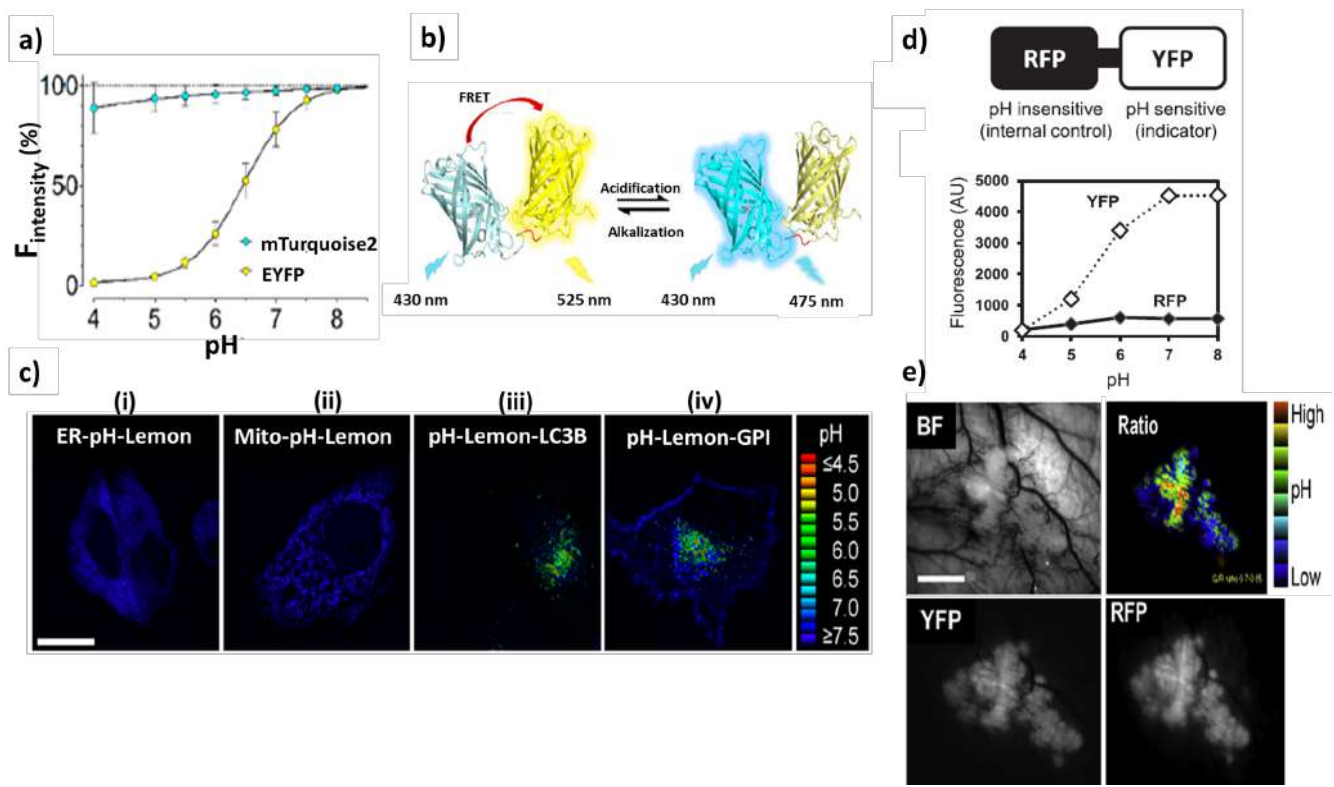

**Figure 4 – Use of different fluorescent proteins for pH imaging**. **(a)** Fluorescence intensities of mTurquoise2 and EYFP at different pH values; **(b)** FRET-based sensing mechanism of pH-Lemon; **(c)** Expression of pH-Lemon in HeLa cells targeted to (i) the ER lumen, (ii) the mitochondrial matrix, (iii) autophagosomes and autophagolysosomes, or (iv) a GPI-anchor (Scale bar: 10 μm);[192] **(d)** Design of RFP-YFP chimeric protein as a ratiometric pH indicator and correlation of its fluorescence intensities with pH; **(e)** Representative images of a tumour xenograft observed with bright field (BF) and fluorescence (YFP and RFP) microscopy.[193]

Additionally, redox dye-based hypoxia-sensitive fluorescent probes have been designed in which the functional groups, such as an azo group, can be cleaved under hypoxic conditions. Based on this, *Kiyose* et al. developed a series of NIR 'turn-on' fluorescent probes, QCys (NIR dyes cyanine and quencher BHQ linked together by azo functional groups) for imaging of hypoxic cells and real-time monitoring of ischemia in the liver and kidney of live mice **(Figures 5c and 5d)**.[208] As these turn-on probes rely on a series of chemical reactions, irreversibility of the sensors could pose a major limitation during real-time sensing of hypoxia, where the oxygen concentration may change overtime. Using a similar strategy, *Sun* et al. reported azo-modified Ir(III) complexes to detect hypoxia in 3D multicellular tumour spheroids, which produced phosphorescence after reduction by azo-reductases under hypoxic conditions.[209] These probes can penetrate over 100 μm into 3D multicellular spheroids to reach the hypoxic regions. However, in this work too, irreversible unidirectional detection could be a limitation in real-time dissolved oxygen fluctuation monitoring.

In general, imaging based on redox-sensitive Nitroimidazole derivatives is efficient and straightforward and can be performed in cells and animals using a basic fluorescence microscope. However, it is an indirect method (non-quantitative) and may have a high background signal due to the instability of these probes *in vivo*.[103,159]

*b) Nanosensors:* Several nanoparticle-mediated methods have been employed for fluorescence-based detection of cancer biomarkers which offer properties such as high surface area-to-volume ratio and unique optical properties.[210–212] *Wu* et al. have described multifunctional hypoxia imaging nanoparticles (core-shell structure nanoparticles, matrix-dispersed nanoparticles, self-assembled nanoparticles, and micelle/liposome-like nanoparticles) for tumour imaging and guided tumour therapy.[213] Recently, *Zheng* et al. reported a NIR phosphorescent ratiometric nanoprobe composed of a fluorescent semiconducting polymer and a Ppalladium complex, which enables quantitative imaging of tumour hypoxia dynamics during radiotherapy.[214] Another example of an NIR-excitable oxygen probe was reported by *Lv et al.* describing a modified phosphorescent iridium(III) complex for hypoxia bioimaging via up- and down-conversion luminescence time-resolved luminescence microscopy.[215] *Liu et al.* have described ultrasensitive nanosensors based on upconversion nanoparticles for selective hypoxia imaging *in vivo* upon NIR excitation using [Ru(dpp)₃]Cl₂ as oxygen indicator.[216] The reversible probe composed of two moieties ([Ru(dpp)₃]Cl₂ and up-conversion nanoparticles) becomes luminescent under hypoxic conditions and vice versa. This platform was utilized to image hypoxic regions with high penetration depth in cells and zebrafish. Similarly, several other up-conversion nanoparticles have been used for hypoxia imaging in tumour tissue.[217,218]

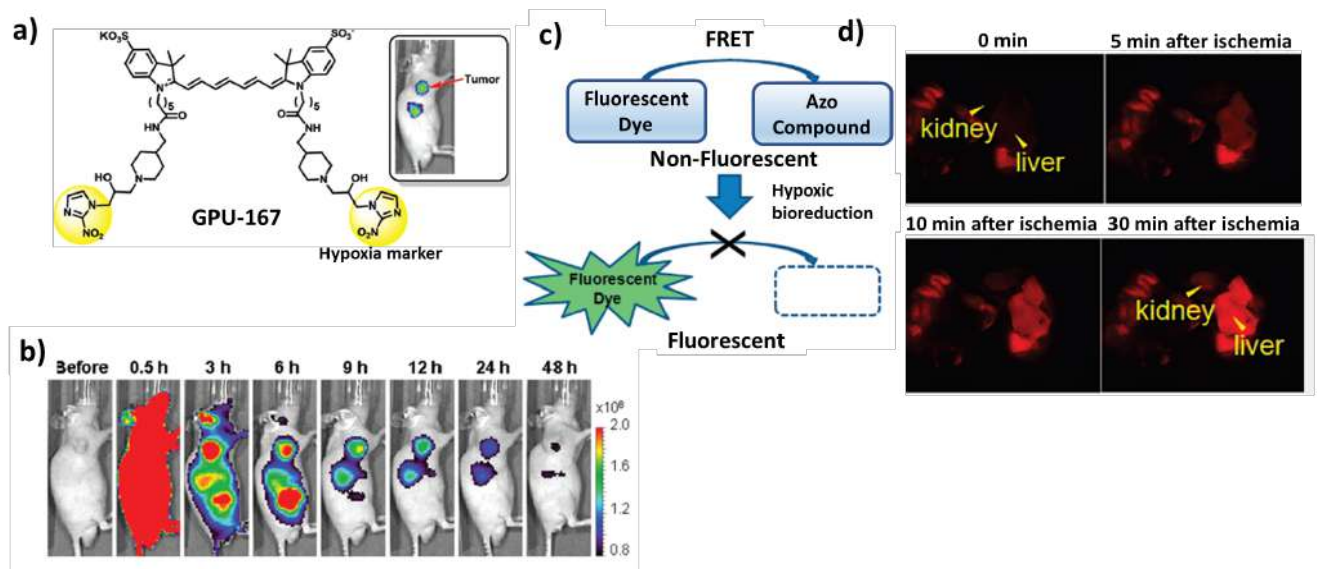

**Figure 5 – Use of bio-reductive probes for hypoxia imaging. (a)** Chemical structure of GPU-167; **(b)** *In vivo* optical imaging with GPU-167 in nude mice carrying SUIT-2/HRE-Luc xenografts;[203] **(c)** Working mechanism of QCys; **(d)** Fluorescence images after injection of QCy5 in living mouse followed by vessel ligation at different time points.[208]

An example of a targeted oxygen sensor was reported by *Napp et al.* who use luminescent NIR polymer-nanoprobes for *in vivo* imaging of tumour hypoxia.[219] They developed a ratiometric probe, where polystyrene nanoparticles (PS-NPs) were doped with oxygen-sensitive Pd-palladium Meso-tetraphenylporphyrin and with an inert reference dye, both excitable at 635 nm **(Figure 6a)**. Dual-wavelength and lifetime-based photoluminescent measurements were employed for oxygen sensing. In *in vitro* and *in vivo* experiments, they targeted trastuzumab-conjugated nanoparticles to cells overexpressing HER2/*neu*, demonstrating the successful use of targeted NIR-based nanosensors in the imaging of hypoxia **(Figure 6b)**.

*Palmer et al.* have described techniques by which vascular and tissue oxygenation can be imaged at high spatial and temporal resolution. They used hyperspectral imaging to characterize haemoglobin absorption to quantify haemoglobin content and oxygen saturation, while dual emissive fluorescent/phosphorescent boron nanoparticles served as ratiometric indicators of tissue oxygen tension.[220] Recently, *Zheng* and colleagues synthesized a water soluble phosphorescent Ir-PVP (poly(N-vinylpyrrolidone)) NIR nanoprobe specific to the hypoxic TME.[221] This probe was capable of detecting hypoxia in metastatic tumours *in vivo* as well as identifying cancer cells at a very early stage of tumour development. The detection was based on the increased oxygen consumption by proliferating cancer cells which can cause hypoxia **(Figures 6c-e)**.

*c) Metal-ligand complexes and metalloporphyrins:* Probes based on metal-ligand complexes have long lifetimes, absorb in the visible region, have significant Stokes shift and are photostable all traits that make them suitable for imaging applications.[204] Probes based on Ru-ruthenium (Ru) and Ir-iridium (Ir) metal complexes are the most commonly used for real-time imaging of hypoxia. They rely on phosphorescence quenching by oxygen due to energy transfer between triplet states of oxygen and the metal complex which emit phosphorescence where the oxygen supply is insufficient.[222] Phosphorescent Ru-ruthenium complexes modified with a Nitroimidazole group that can image oxygen fluctuation in real-time in both living cells and murine tumours in living cells and tumour tissue planted in mice) have been described by *Son et al.*[223] The simple modification of the complex resulted in visualization of tumour tissues and their oxygen fluctuation. The incorporation of Nitroimidazole units into water-soluble Ru-ruthenium complexes made them permeable to cells, allowing their accumulation in hypoxic cells *in vivo* (**Figures 7a-c**).

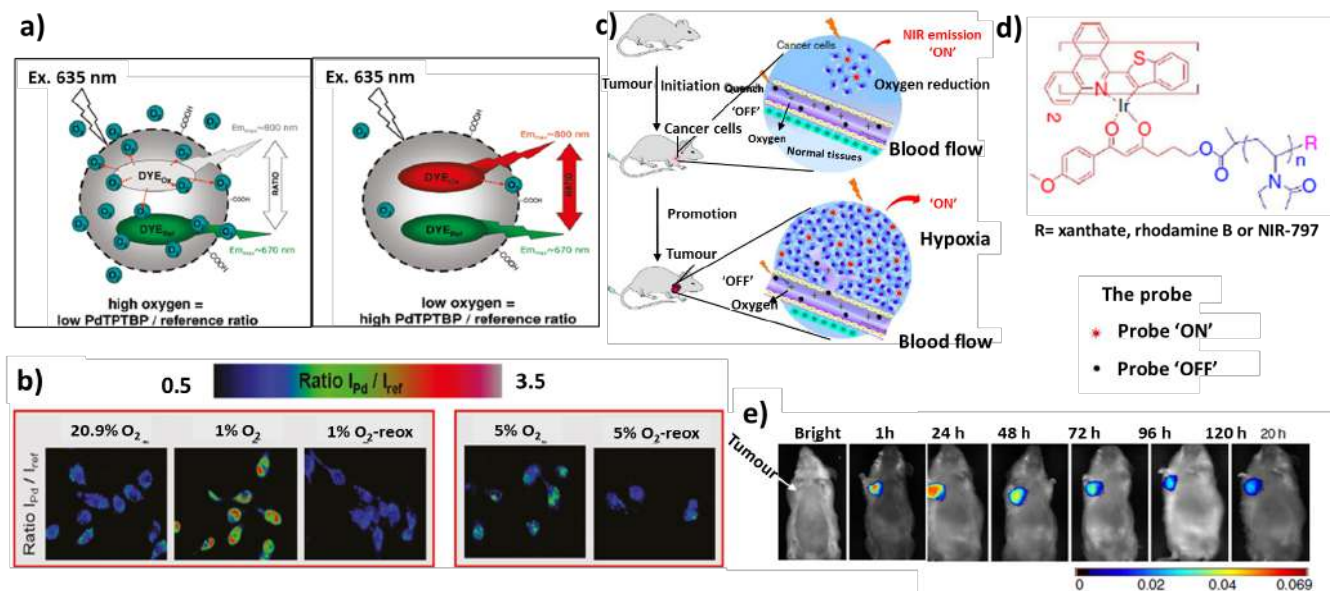

**Figure 6 – Use of different nano-sensors for hypoxia imaging**. **(a)** Schematic representation of hypoxia sensing mechanism of PS-NPs doped with an oxygen-sensitive Pd palladium Meso-tetraphenylporphyrin and an inert reference dye; **(b)** Ratiometric response of the probe in MH-S cells under different oxygenation conditions; [219] **(c)** Schematic of tumour/cancer cell detection with the Ir-PVP probe; **(d)** Chemical structure of the Ir-PVP probe; **(e)** Optical imaging with the Ir-PVP probe in a ICR mice bearing H22 tumour at different time points.[221]

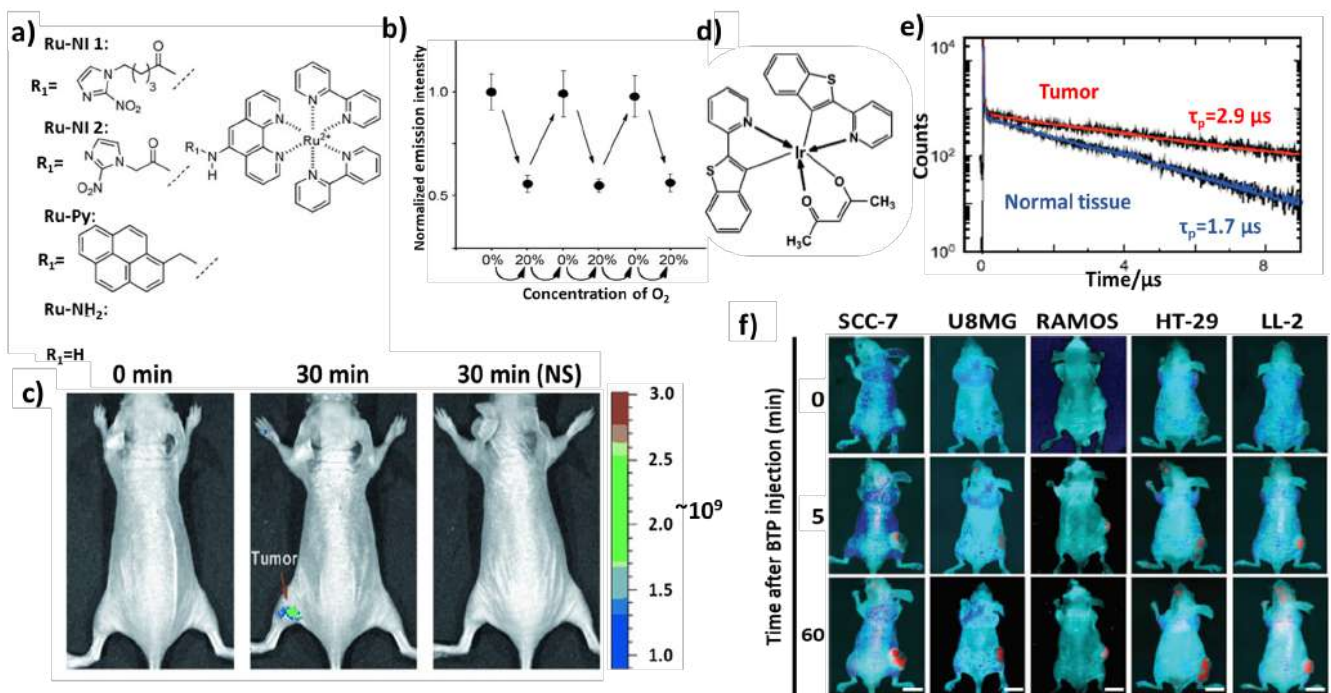

**Figure 7 – Metal-ligand complexes as hypoxia imaging agents**. **(a)** Molecular structures of Ru ruthenium complexes; **(b)** Reversibility in phosphorescence of Ru-NI 1 at 595 nm to alternating changes of oxygen concentration; **(c)** Optical imaging of tumour hypoxia in tumour-bearing nude mice after the injection of Ru-NI 1 or normal saline (NS) at different time points[223]; **(d)** Chemical structure of BTP; **(e)** Comparison of decay profiles for tumour moiety and normal tissue monitored at 620 nm; **(f)** *In vivo* optical imaging of tumour-bearing athymic nude mice with BTP at different time points (Scale bars: 10 mm).[224]

Iridium-based metal complexes have also been used for imaging hypoxia in tumours. For example, *Sun et al.* used Azo-based Iridium (III) complexes as multicolour phosphorescent probes to detect hypoxia in 3D tumour spheroids.[209] *Zeng et al.* developed a phosphorescence imaging approach using BTP [Bis(2-(2'-benzothienyl)-pyridinato-N,C³')Iridium (acetylacetonate)] to monitor hypoxic changes resulting from the chemotherapeutic effects of cisplatin *in vivo*.[40] This BTP complex has been utilized by *Zhang et al*.,[224] as a hypoxia sensing probe for tumour imaging *in vivo*. Using phosphorescence lifetime measurements, the authors demonstrated the localization of BTP in the tumour cells. Additionally, they were able to extend the penetration limit of the BTP probe by modifying its acetylacetone moiety **(Figures 7d-f)**.

Metalloporphyrins complexed with Pplatinum(II), Pppalladium(II), and Zzinc(II), are the most commonly used optical oxygen sensor devices as their phosphorescence is strongly quenched by oxygen.[225–227] Recently, *Fang et al.* reported compact conjugated polymer dots (Pdots) with covalently incorporated metalloporphyrins for bioimaging of hypoxia.[228] PFPtTFPP, a kind of hydrophobic polymer, was used as an imaging probe with fluorene as an energy donor and an oxygen-sensitive PtII porphyrin as an energy acceptor. *Zhao et al.* also reported the use of hydrophilic phosphorescent starburst Pt(II) porphyrins as bifunctional therapeutic agents for tumour hypoxia imaging and photodynamic therapy (PDT).[229] They synthesized a series of 3D phosphorescent starburst Pt(II) porphyrins by using Pt(II) porphyrins as the functional core and cationic oligofluorenes as the arms. The probes had a high signal-to-noise ratio and ultrasensitive oxygen-sensing performance with high quantum yields. The authors also evaluated the PDT effects, both *in vitro* (under hypoxic and normoxic conditions) and *in vivo*.

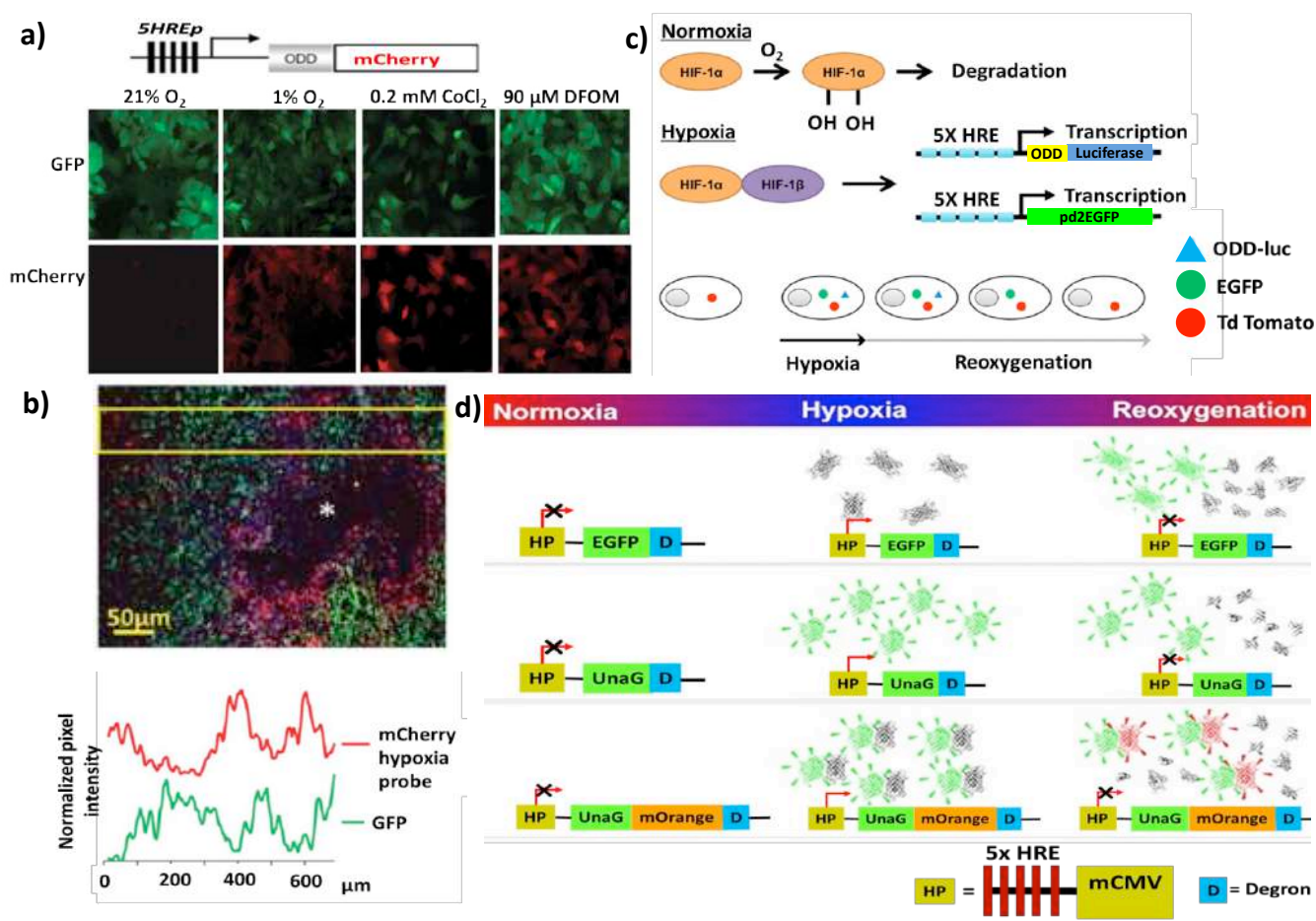

**Figure 8 – Use of fluorescent proteins for hypoxia imaging**. **(a)** Schematic diagram of 5HRE-ODD-mCherry construct and images of GFP MET1-5HRE-ODD-mCherry cells under normoxic and hypoxic conditions; **(b)** Image of GFP MDA-MB-231-5HREODD-mCherry derived frozen tumour sections and quantification of normalized GFP or mCherry signal;[230] **(c)** Schematic description of the "timer" strategy for imaging HIF in prostate cancer cells;[231] **(d)** Schematic representation of hypoxia detection mechanism of UnaG probes.[232]

Metal-based probes exhibit oxygen-dependent signals; thus, quantitative hypoxia imaging with high specificity can be achieved. However, the toxicity of these probes due to the use of heavy metals is a significant concern. Consequently, strategies for decreasing the toxicity of these fluorescent probes could prove beneficial for hypoxia imaging in future.[103]

*d) Genetically encoded probes:* Probes based on oxygen-sensitive fluorescent proteins such as GFP, YFP or DsRed, are biocompatible, allow intracellular targeting, and exhibit high brightness and tunable spectral characteristics.[159]

*Takahashi et al.* utilized the phenomenon of the redshift of GFP fluorescence upon photoactivation for *in vivo* molecular imaging of oxygen.[233,234] Recently, the ratiometric biosensor DsRed FT (Timer), which is based on different $O_2$-dependent maturation rates of isoforms of DsRed protein, was described by *Lidsky et al.* Ratiometric analysis of green (low $O_2$) and red (high $O_2$) fluorescence allows simple analysis of hypoxia in tissue and whole animals.[235]

Most of the fluorescent protein reporter-based systems are controlled by a hypoxia-responsive promoter such as 5HRE, which is an artificial HIF-1-dependent promoter that can cause 100 times more expression of luciferase in xenograft.[236] In another related study, *He et al.* developed a novel xenograft model for studying and imaging hypoxia-induced gene expression using a hypoxia-inducible dual reporter under the control of the hypoxia response element (9HRE).[237] *Wang et al.* reported a novel fluorescent mCherry hypoxia-responsive marker that can be used for real-time imaging of hypoxic cells at single cell resolution**(Figures 8a and 8b)**.[230] *Danhier et al.* combined optical reporter proteins with different half-lives to detect temporal evolution of hypoxia and reoxygenation in tumours ("timer" strategy).[231] They used a luciferase gene fused with an oxygen-dependent degradation domain (ODD-luc) and a EGFP gene, both of which were under the control of a 5xHRE promotor **(Figure 8c)**. *Erapaneedi et al.* described a novel family of genetically encoded fluorescent sensors which revealed substantial heterogeneity in tumour hypoxia at the cellular level.[232] The sensors are based on a HIF-inducible promoter reported by the oxygen-independent fluorescent protein UnaG (Japanese eel green fluorescent protein). Also, the authors developed a fusion sensor comprised of UnaG and mOrange, which could identify reoxygenated cells, in addition to hypoxic cells. The sensor incorporates different switching and memory behaviours and allows visualization of tissue hypoxia with light microscopy **(Figure 8d)**.

### 3.3 Optical Metabolic Imaging

Optical metabolic imaging (OMI) probes the autofluorescence intensity and lifetime of the two metabolic coenzymes NAD(P)H and FAD for robust detection of metabolic states at cellular resolution. OMI is well suited for evaluating heterogeneous drug responses due to its single-cell resolution and high sensitivity to changes in cellular metabolism.[238] NADH and FAD each have two-component fluorescence decays. For NADH, the short lifetime ($\tau_1$) corresponds to NADH free in solution, whereas the long lifetime ($\tau_2$) corresponds to protein-bound NADH.[239] Conversely, protein-bound FAD corresponds to the shorter lifetime, whereas free FAD corresponds with the long lifetime.[240] The shorter fluorescence lifetimes of both protein-bound FAD and free NADH are due to dynamic quenching by their adenine moiety. The mean fluorescence lifetime ($\tau_m$) is the weighted average of the short and long lifetime components, $\tau_m = \alpha_1\tau_1 + \alpha_2\tau_2$, where $\alpha_1$ and $\alpha_2$ are the fractional contributions of the short and long lifetimes, respectively.

As OMI is inexpensive, less time consuming, and directly assesses dynamic variation in cellular metabolism, it can be used to assess glycolysis, [241] oxidative phosphorylation,[242] and metabolic enzyme microenvironment interactions.[243] Most of the studies on OMI involve the use of custom-built multiphoton fluorescence microscopy systems for fluorescence imaging and time correlated single photon counting to acquire the lifetime images, where each pixel of the image indicates lifetime. Using OMI, single cell resolution allows for identification of pre-existent resistant subpopulations of cells that can cause relapse after the therapy.

Initially, the optical redox ratio and the ratio of lifetimes were correlated with drug treatments. Later, various groups devised index values by combining redox ratios and lifetimes in different ways to generate a single quantifiable parameter that can reflect the metabolic profile of cells. One such example is the OMI index,[244] which is the linear combination of optical redox ratio and fluorescence lifetimes of NADH and FAD in living cells and tissues. The OMI index was then correlated with metabolic changes in correlation with drug response.

*Walsh et al.* used live hamster cheek pouch epithelia tissue culture experiments to serve as a surrogate for *in vivo* metabolic measurements.[245] The optical redox ratio (fluorescence intensity of NADH/FAD) of the cultured biopsies was statistically identical to the *in vivo* measurements for 24 h, while the redox ratio of the frozen-thawed samples decreased by 15% ($p < 0.01$). The NADH mean fluorescence lifetime ($\tau_m$) remained unchanged ($p > 0.05$) during the first 8 h of tissue culture, while the NADH $\tau_m$ of frozen-thawed samples increased by 13% ($p < 0.001$). Notably, the cellular morphology did not significantly change between *in vivo*, cultured, and frozen-thawed tissues ($p > 0.05$) **(Figure 9)**. This study suggests that for optical metabolic and morphological measurements, short-term *ex vivo* tissue culture may be more appropriate to recapitulate the *in vivo* status than frozen-thawed tissue.

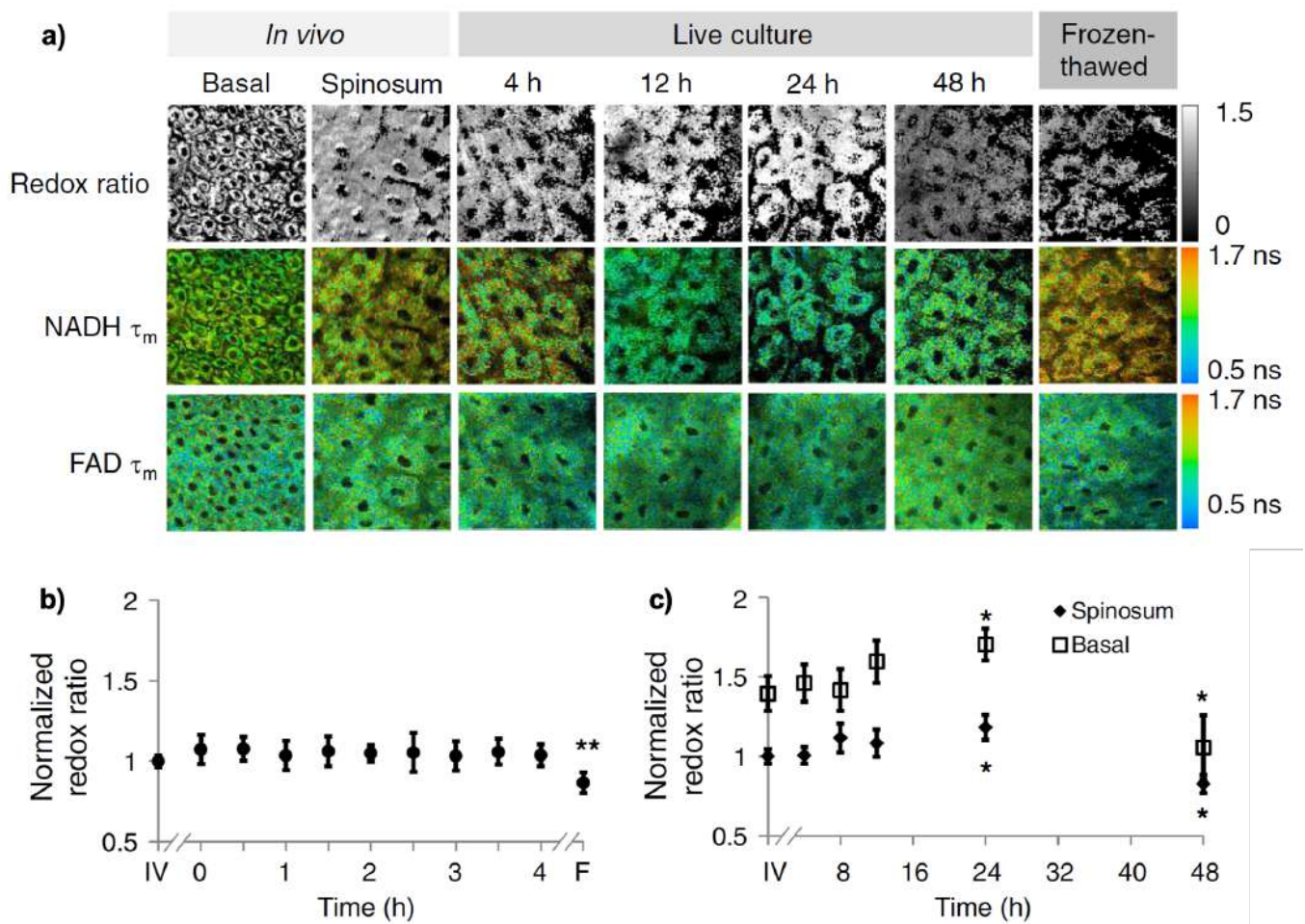

**Figure 9 – (a)** Representative redox ratio (first row), NADH $\tau_m$, and FAD $\tau_m$ images from *in vivo* hamster epithelium. $\tau_m$ is the mean lifetime ($\tau_1 * \alpha_1 + \tau_2 * \alpha_2$); **(b)** The redox ratio value of frozen-thawed (F) hamster epithelium is significantly reduced from the *in vivo* (IV) value, while live tissue culture showed similar ratios similar to *in vivo* values during the initial 4h of culture; **(c)** The redox ratios of the spinosum and basal epithelial layers of the live-culture biopsies after 48 h (mean ± SE of n = 6).[245]

*Walsh et al.* later investigated the association between glycolytic levels, tumour subtypes, and early-treatment response in breast cancer using OMI.[241] They reported that the OMI index could be correlated with glycolytic levels across a panel of human breast cell lines, using standard assays of cellular rates of glucose uptake and lactate secretion. OMI was successfully used to resolve differences in the basal metabolic activity of untransformed versus malignant breast cells ($P < 0.05$) and between breast cancer subtypes ($P < 0.05$), defined by the expression or absence of the estrogen receptor and/or HER2. They also assessed metabolic changes induced by trastuzumab in HER2-overexpressing human breast cancer xenografts in mice. The OMI was significantly faster in capturing the drug response compared to fluorodeoxyglucose-positron emission tomography (FDG-PET). Inter-cellular heterogeneity, in terms of drug response, was also observed, indicating cellular subpopulations that constitute a significant cause of cancer relapse.

Similarly, *Shah et al.* studied the treatment response of human head and neck squamous cell carcinoma (HNSCC) using OMI.[246] In this work, two HNSCC cell lines, SCC25 and SCC61, were treated with cetuximab (anti-EGFR antibody), BGT226 (PI3K/mTOR inhibitor), or cisplatin (chemotherapy) for 24 hours. The shift in early metabolic changes induced by drug treatment was accurately measured by OMI. Later in another study, the authors performed successful metabolic imaging of head and neck cancer organoids after treatment with standard therapies, including an antibody therapy, chemotherapy, and combination therapy.[247] They were also able to analyse organoid cellular heterogeneity both quantitatively and qualitatively. Their results indicated that OMI is sensitive enough to capture the therapeutic response in organoids after just 1 day of treatment and, can also resolve cell subpopulations with distinct metabolic phenotypes. This platform could be thus used for streamlining the drug discovery process for head and neck cancers.

Importantly, in 2014 *Walsh et al.* reported using OMI to effectively predict the drug response in breast cancer.[244] As soon as 24 hours post-treatment with clinically relevant anticancer drugs, the OMI index of responsive organoids decreased and, was further reduced after effective therapies were combined. Drug-resistant organoids showed no change. Heterogeneous cellular responses to drug treatment were also resolved in organoids using OMI. The *in vitro* drug response data were validated *in vivo* in xenograft tumours derived from organoids with tumour growth measurements and staining for proliferation and apoptosis. As an example of OMI application in pancreatic cancer tumoroids, *Walsh et al.* investigated murine and human pancreatic cancer organoids, revealing a heterogeneous drug response.[242] In comparison to previous OMI-related studies, the authors also used an OMI index to correlate metabolic changes directly with drug response.[248]

The OMI index allowed the comparison of metabolic states across experimental groups, where a significant decrease in the OMI index indicates a positive drug response. Using this approach, the authors found differential drug responses in different subpopulations of tumour cells and tumour-associated fibroblasts within the same cultures. OMI was also employed to study the effect of treatment with either the antibody cetuximab or the chemotherapeutic cisplatin on mice with FaDu (human hypopharyngeal carcinoma cell line) xenograft tumors.[249] Interestingly, the *in vivo* OMI results revealed the effect of treatment within only in two days, compared to standard assessment of tumour size through palpation and caliper measurements, which took nine days. Additionally, the authors developed a dimensionality reduction technique (viSNE), which enabled holistic visualization of multivariate optical measures of cellular heterogeneity **(Figure 10)**. Overall, their study allowed visualization of increased heterogeneity in cetuximab and cisplatin treatment groups compared with the control group.

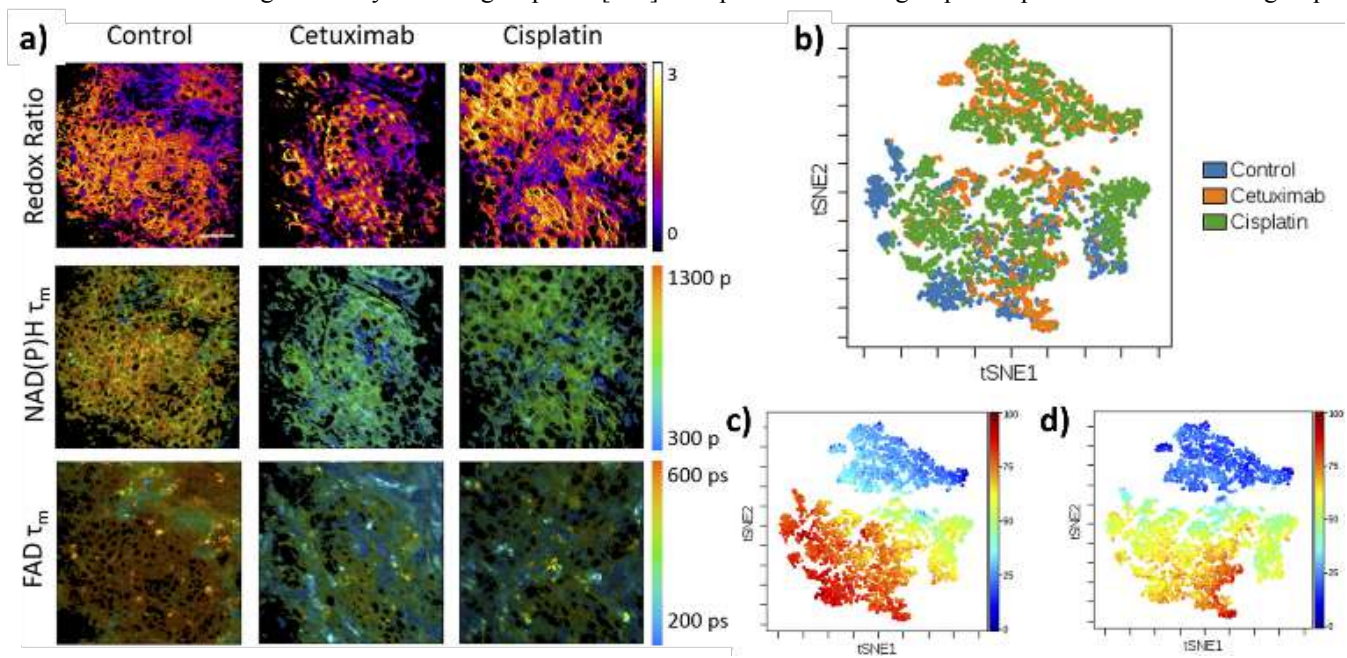

**Figure 10 – (a)** *In vivo* images of FaDu xenografts 2 days after treatment with cetuximab or cisplatin. NAD(P)H and FAD fluorescence images were selected from the same field, and the redox ratio (top row), NAD(P)H fluorescence lifetime (middle row), and FAD fluorescence lifetime (bottom row) were calculated. Scale bar = 50 µm; **(b)** Single-cell analysis using the viSNE results in two dimensions for visualization of heterogeneity across individual cells. The cetuximab and cisplatin treatment groups show some overlap with the control group, but also exhibit a separate subpopulation of cells. Heat maps of the short fluorescence lifetime components for **(c)** NAD(P)H and **(d)** FAD, show gradients over the 2-dimensional viSNE axes.[249]

## 4. Magnetic Resonance Imaging of the Tumour Microenvironment

The microenvironment of tumours is distinct from that of normal tissues, as mentioned above, due primarily to alterations in oxygen partial pressure, in pH levels, glycolytic activity, in the interstitial fluid pressure and the vascular network, in addition to the abnormal behaviours exhibited by the tumour and immune cell components. These conditions create a tumour-specific homeostasis that can promote the proliferative and invasive capabilities of cancer cells, resulting in a more aggressive disease phenotype and consequent enhanced resistance to treatment. Alterations in the TME can also be exploited to study and characterize the disease through well established, non-invasive *in vivo* imaging modalities such as MRI, CT and PET.

Decades ago, the advent of proton-based MRI enabled detailed images of the human anatomy that far surpassed those seen using radiological imaging techniques. In principle, MRI is based on the measurement of radio frequency signals released by atoms after they have absorbed radio frequency energy in an external magnetic field. The imaging of the protons of hydrogen atoms, which are highly abundant in the water, blood, and fat of living organisms, is the basis of most MRI. The differential amount of water in tissues acts as a natural contrast agent (CA) that represents their intrinsic biological properties. Tissues are most commonly visualized via techniques referred to as T1- or T2-imaging, both of which measure the chemical properties of water (protons) in tissue. T1 images display water as being hypointense (dark) while, in T2 images, water is hyperintense (bright); therefore, MRI is broadly based on 'T1-weighted' and 'T2-weighted' sequences (sets of radiofrequency pulses and gradients).

Since the early work of *Damadian*, *Lauterbur* and *Mansfield*, MRI has been viewed as an unlikely candidate for imaging the TME, due to its limited imaging spatial resolution in low field magnets. However, the advent of ultra-high field scanners with ever-stronger field strengths, in both the clinical and preclinical settings, has fundamentally changed the imaging capabilities of MR. Significantly enhanced resolution, down to 500 microns per axis in a 7 tesla or, 80 microns in a 11.7 tesla scanner is now possible.[250] Greater signal-to-noise ratios and faster imaging speeds, in addition to the development and refinement of new MR techniques, now permits a wide range of applications, making MRI one of the most versatile imaging modalities. Unlike other modalities such as CT and PET, MRI is not based on ionizing radiation, allowing safe and repetitive imaging for longitudinal studies. With no limitations in the depth of penetration (as is the case with optical imaging) or in tissue visualization, unlike optical imaging, MRI provides extremely clear, detailed anatomical images. In addition, MRI supports functional, diffusion, and perfusion imaging (described below) and can be combined with other imaging modalities, to generate complementary information. A multitude of novel and specialized MRI sequences have been developed that have the capacity to simultaneously provide information maps of a host of parameters such as tumour $O_2$, metabolism and vascularization, greatly contributing to our understanding of both the healthy and the TME.

While MRI can measure the signal generated from endogenous protons, it is also easily amenable to the use of exogenous CAs to provide a wider dynamic range and enhanced specificity. The MRI CA field continues to evolve, generating ever more sophisticated molecules and nano-agents, ranging from simple well-established molecules to complex multi-functional compounds with diagnostic as well as therapeutic capabilities.

With all the major advances in MRI, a detailed account of all applications cannot be encapsulated in this review; therefore, the most widely used techniques, along with promising applications for the study of the TME will be presented. Lastly, the versatility of MRI sometimes enables the evaluation of a multitude of parameters by one single technique; for example, the profiling of pH, lactate, or glycolytic/oxygen metabolism can be all performed using chemical exchange saturation transfer (CEST) MRI, making it difficult to establish strict categorizations based on their applications. **Table 2** summarizes the major MRI applications.

## 4.1 Magnetic Resonance Imaging of Acidosis

A hallmark of the TME is its altered pH, a result of abnormal metabolism, hypoxia, and poor vascularization. An abnormal pH can in turn, affect the behaviour of tumours, their progression and their response to treatment.

Chemical exchange saturation transfer MRI (CEST-MRI) is an innovative technique that enables the imaging of compounds found in very low concentrations with a high spatial resolution and sensitivity. CEST-MRI takes advantage of the biophysical properties of biological compounds, where magnetization in the form of exchangeable protons is transferred from diverse molecules to water. After excitation with a specified radio frequency pulse, the state of low magnetization and signal reduction (saturation) that was originally on the targeted species is transferred to water. This decrease in signal is easily detectable using standard MRI sequences, and is proportional to the concentration of the specific molecule. CEST MRI can utilize either endogenous or exogenous compounds and, provides semi-quantitative mapping of acidosis and metabolism through the imaging of compounds such as proteins, peptides and sugars.[251]

CEST-MRI terminology is typically based on the specific functional group detected at a certain frequency shift. For example, amide proton transfer (APT)-CEST-MRI can detect the protons of amide groups (-NH) of endogenous myelin proteins and peptides that resonate at ~3.5 ppm, and whose localized biochemical activity is altered in many conditions such as tumours and ischemic strokes. Amine-CEST detects exchangeable protons in molecules with amine groups (-$NH_2$), which resonate at ~1.8–3.0 ppm, on metabolites such as creatine and glutamate. Molecules with hydroxyl groups (-OH), resonating at ~0.5–1.5 ppm, such as those found in glycosaminoglycan, myo-inositol, and glucose, are imaged with hydroxyl CEST.[252] CEST-MRI can also be named according to whether it evaluates acidity (acidoCEST), detects lactate (LATEST) or whether it focuses on the metabolism of glutamate (GluCEST). These modalities will be discussed again in section 4.3.

**Table 2.** *MRI techniques to study the tumour microenvironment*

| | Acidosis | Hypoxia | Metabolism | Vasculature | Other |
|---|---|---|---|---|---|
| ***CEST-MRI*** | Intra- and extracellular pH (PARACEST CAs), lactate | Tissue redox state[253] oxygen consumption[254,255] | Lactate, glucose, glutamate, creatine, glycogen and glycosaminoglycan | Tissue perfusion[256–258] BBB permeability[258] | Necrosis, oedema Tumour diagnosis; monitoring tumour progression and response; predict treatment response[259] |
| ***AACID*** | Amine: amide ratio | | | | Monitoring tumour response |
| ***$^{13}C$ Hyper-polarized MRI*** | $^{13}C$-bicarbonate for pH | $^{13}C$-dehydroascorbate and $^{13}C$- acetoacetate for redox status. | $^{13}C$-pyruvate for glycolysis, mitochondrial energy production and amino acid synthesis | $^{13}C$-urea for perfusion | $^{13}C$-fumarate for necrosis. $^{13}$lactate: $^{13}$bicarbonate ratio for tumour response |
| ***BOLD-MRI*** | | Blood and tissue oxygenation | Oxygen consumption, blood and tissue oxygenation | | Tumour grading, tumour response to treatment. |
| ***TOLD-MRI*** | | Tissue oxygenation, tumour pO2 | | | Tumour grading, tumour response to treatment. |
| ***EPRI*** | | Molecular oxygen, quantitative maps of pO2, tumour heterogeneity, chronic and cycling hypoxia[260] | | Blood flow, blood volume[261] | Prediction of tumour response to treatment |
| ***DCE-MRI*** | | Tissue oxygenation and perfusion | | Tissue perfusion, blood volume, vascular permeability, BBB integrity, fluid volume fractions | Tumour diagnosis and grading, tumour response to treatment. |
| ***DSC-MRI*** | | | | Tissue perfusion, blood flow, blood volume, BBB permeability | Tumour response to treatment. |
| ***MRI Spectroscopy*** | Lactate | | Metabolism of glutamate, choline, creatine, citrate, myo-inositol, taurine, lactate, lipids. | | Tumour diagnosis and grading, tumour response to treatment. |
| ***ASL-MRI*** | | Tissue hypoxia | | Blood flow, arterial transit time, tissue perfusion | Tumour diagnosis and grading; response to anti-angiogenics |
| ***DWI, DTI*** | | Tissue perfusion | | Tissue perfusion | Microstructure,[262] fibrosis,[263] tumour grading,[264]prognosis |

*a) Intracellular pH (pHi):* Intracellular pH (pH$_i$) varies in many disease states, including cancer, affecting many biological processes such as enzymatic reaction rates, energy metabolism and reactive oxygen species production. Therefore, pH$_i$ is a known biomarker of altered cellular function and of response to interventions such as radiation therapy.[115] MRI is uniquely capable of imaging the pH$_i$ through several different methods.

Amine and amide concentration-independent detection (AAICD) is a ratiometric approach that takes advantage of the differential sensitivities of amine and amide CEST to pH changes. Thus, the distinct CEST readouts on these two functional groups enables the generation of quantitative pH maps.[265] Since 90% of protein in tissue exists predominantly in the intracellular space, this measurement is often considered **pH$_i$**, but technically, it measures both intra- and extra-cellular tissue. This approach was applied to study the response of glioblastoma multiforme (GBM) in an orthotopic U87 cell mouse model to topiramate (TPM), an inhibitor of carbonic anhydrase. Using a 9.4 T MRI scanner, *Marathe et al*. observed that TPM caused a marked increase in intracellular acidosis of implanted tumours.[266] pH$_i$ can also be non-invasively imaged with amine-CEST, based on the pH sensitivity of the amine protons of glutamate and glutamine; the shortcoming of this technique is that it is also sensitive to the concentration of these protons. *McVicar et al*. examined the effect of lonidamine, (an inhibitor of lactate transport that acidifies the intracellular space) using a similar mouse model of GBM as above, and found that the effectiveness of APT-CEST and AACID in detecting pH changes were comparable in sensitivity at 9.4 T.[267]

It is also possible to measure pH$_i$ non-invasively using $^{31}$P (rather than $^1$H/proton) MR spectroscopy, by comparing the chemical shift from the pH-dependent inorganic phosphate peak and the pH-independent peaks of either phosphocreatine or alpha adeno triphosphate.[268] Spectroscopy will be discussed in detail in section 4.3.

*b) Extracellular pH (pHe):* While MRI measurements of pH$_i$ primarily use endogenous molecular groups as CAs, the imaging of pH$_e$ by CEST-MRI, predominantly uses exogenously administered pH reporters called paramagnetic CEST (PARACEST) CAs. These CAs contain a lanthanide ion and are normally restricted to the extracellular space, and significantly shift the MR frequency of the exchangeable protons away from the frequency of water, thus providing a wide dynamic range.[269]

For example, *Sheth et al*. measured pH *in vitro* using the PARACEST MRI lanthanide CA, Yb-DO3A-oAA. Interestingly, this CA exhibits two CEST effects within the same compound that are dependent on pH[270] allowing precise ratiometric analyses of pH that is independent of concentration and T1 relaxation times, but that is sensitive to temperature. *Castelli et al* used a similar compound, Yb-HPDO3A, to measure pH at different stages of tumour development of a subcutaneous mouse model of melanoma.[271]

Ever since *Aime et al*. began exploring the use of iodinated low-osmolar radiocontrast agents commonly used for CT-based diagnosis in CEST MRI,[272] these CAs have gained wide appeal as they exhibit relatively low toxicity and their multi-site pH-dependent chemical exchange properties are ideal for ratiometric measurements of pH. *Chen et al*. used Iopramide for acidoCEST MRI in a MDA-MB-231 mouse model of breast cancer to successfully measure pH$_e$ in the tumour, kidney and bladder (pH range 6.2-7.2) and acidoCEST was highly effective in detecting an increase of tumour pH following bicarbonate alkalinization treatment.[273] Another iodinated CA, Iopamidol, was tested by *Sun et al*. to assess pH between 6.0 and 7.5 *in vitro*[274] and by *Moon et al*.[275] *in vivo* in mice with either MCF-7 breast cancer or Raji lymphoma xenograft tumours, concluding that Iopramide and Iopamidol exhibit a similar performance for pH measurements. Yet another ratiometric iodinated CA, Ioversol, was used to distinguish between tumour and normal tissue in a rat hepatoma model in the pH range of 6.0 to 7.8.[276] Refinements in both the MR sequences and in the analysis of this technique for the spatial mapping of pH$_e$ continue to be made.

*c) Lactate Imaging:* In the hypoxic environment of tumours, cancer cells have a preference for anaerobic glycolysis due to numerous abnormalities in the activity of metabolic enzymes such as lactate dehydrogenase (LDH), leading to the increased production of lactate and consequent acidification of the TME. The two main MRI methods to image lactate are magnetic resonance spectroscopy (see section 4.3.) and CEST–MRI of lactate (LATEST), which measures the exchange saturation between the lactate hydroxyl proton and water protons. *DeBrosse et al*. used LATEST to map dynamic changes in lactate metabolism in a murine lymphoma model in response to an injection of pyruvate.[277] There are challenges in applying LATEST, as the hydroxyl protons of lactate resonate very close to those of water and, extracellular lactate content cannot be distinguished from intracellular lactate. To partially address this, *Zhang* used EuDO3A, a lactate-responsive molecule that shifts the lactate -OH resonance away from the tissue water signal, enabling LATEST without interference from other endogenous -OH containing metabolites.[260,278]

## 4. 2 Magnetic Resonance Imaging of Hypoxia

Hypoxia results when the partial pressure of oxygen is too low to meet the demand of the tissue. Hypoxia alters the responses of tumours to chemo- and radiotherapy, decreases anti-tumour immune responses, and increases tumour aggressiveness, leading to worse prognosis. Hypoxia varies both spatially and temporally, reflecting the heterogeneity of the tumour and the cyclical adjustments of the TME to the constantly changing conditions.

*a) Dynamic Contrast Enhanced MRI:* Dynamic contrast enhanced-MRI (DCE-MRI) is a minimally-invasive method in which a T1 CA (often gadolinium-based) is injected intravenously in order to visualize temporal CA passage from the blood vessels into the extracellular space, providing real-time information on the degree of tissue vascularization, vessel permeability, integrity of the blood brain barrier (BBB) (since the CA does not normally cross the BBB), fluid volume fractions[279] and indirectly, of fibrosis. Importantly, DCE-MRI has been shown to correlate directly with tissue oxygenation levels, with high contrast enhancement reflecting a well vascularized tumour, and low levels of enhancement indicating possible fibrosis and a compromised microvasculature. DCE-MRI can therefore quantitatively evaluate the levels of tissue perfusion and hypoxia, providing the means to predict which tumours will be more chemo- or radio-resistant and to monitor therapy-induced microvascular changes, particularly in response to anti-angiogenic drugs and radiotherapy. DCE-MRI has been successfully used to characterize the TME of numerous cancers and evaluate treatments of cervical,[280,281] pancreatic,[282–285] and breast[286–288] cancers, among others, both in preclinical and clinical settings. Limitations of DCE-MRI include its relative non-specificity and the lack of methodological standardization between different imaging facilities, making reproducibility of results challenging at times. However, its sensitivity can be increased when combined with other techniques such as BOLD-MRI (discussed next).

*b) Blood-oxygen-level-dependent (BOLD) functional MRI (BOLD-fMRI):* When cellular activity increases, oxygen is extracted from the blood, converting oxyhaemoglobin in red blood cells to deoxyhaemoglobin (a natural endogenous MRI CA). BOLD-fMRI detects the hyperintense signal of deoxyhaemoglobin; however, a quantitative and direct relationship between the BOLD effect and $pO_2$ remains a challenge. BOLD-fMRI exhibits a high sensitivity to oxygen consumption but can be affected by many factors such as blood flow, permeability of the vasculature and haemoglobin levels *per se.*[289] BOLD was first applied to tissue imaging of organs with high energy requirements such as brain,[290] but is now also used to investigate hypoxia and for the assessment of therapeutic efficacy in many types of solid tumours, as reported by *Hoskin et al.* in prostate cancer.[291] *Dallaudiere* also found that BOLD-fMRI was comparable to [18]F-misonidazole PET/CT in identifying hypoxia in osteosarcoma.[292] It is important to note that blood oxygenation does not equal tissue oxygenation;[293] thus, while deoxyhaemoglobin acts as an endogenous T2* BOLD CA, molecular $O_2$ affects T1-weighted imaging, and is the basis of tumour oxygenation level-dependent (TOLD)-MRI. Both these techniques are applied following a gas challenge with either 100% oxygen or carbogen gas, causing arterial hyperoxia. The excess oxygen remains dissolved in blood plasma and interstitial fluid that is detected using a method known as oxygen-enhanced MRI (OE-MRI). For example, *Jerome et al.* tested the capacity of BOLD to identify changes of arterial oxygen and carbon dioxide in response to a respiratory challenge in rat gliomas[294] and *Hallac et al.* used BOLD/TOLD on rat prostate tumours, which correlated with tumour $O_2$ and its response to irradiation.[295] When OE-MRI is combined with dynamic-contrast enhanced perfusion, a method referred to as perfused Oxy-R, hypoxia can be accurately mapped as was shown both in 786–0-R renal cell carcinoma xenografts and in patients with renal cell carcinoma by *O'Connor et al.*[293] Further, in non-small cell lung cancer xenograft models and in patients, Oxy-R MRI showed great promise in the prediction and assessment of tumour response to treatment.[296]

*c) Electron Paramagnetic Resonance Imaging (EPRI):* Electron Paramagnetic Resonance Imaging (EPRI) is a spectroscopic technique in which $pO_2$ changes the relaxation rates of exogenous paramagnetic agents, causing their spectral broadening, which is proportional to the oxygen concentration. Unlike MRI, which detects nuclear spins, EPRI detects unpaired electron spins of free radical species. Upon administration of paramagnetic molecules such as the triarylmethyl radical (TAM) OX063, spatial quantitative maps of $pO_2$ can be generated, allowing the study of its dynamics, its temporal fluctuations in response to anti-cancer therapies, enabling the identification of regions with chronic and cycling hypoxia.[297] *Yasui et al.* used EPRI to evaluate $pO_2$ in two human glioma orthotopic models (with U87 and U251 cells) before and after irradiation; EPRI exhibited heterogeneity of hypoxia within the tumour which showed good correlation with the *ex vivo* staining of the hypoxia probe, pimonidazole.[260]

It is worth mentioning that hypoxia in the TME can also be successfully evaluated by another powerful imaging modality, positron emission tomography (PET). PET is minimally invasive and uses radiotracers such as [[18]F]F-FMISO and Cu-ATSM, which contain oxygen-sensitive Nitroimidazoles that are metabolized differently in aerobic versus anaerobic tissues. As is the case with MRI techniques, PET is selective and sensitive to hypoxic tissue, is quantitative and can help predict therapeutic outcomes. However, PET presents some disadvantages such as a lower spatial resolution and patient exposure to radiation. PET does not permit the assessment of cycling hypoxia and hypoxia-modifying therapies and, importantly, PET does not allow for repeated, longitudinal use. [298–300]

### 4. 3 Magnetic Resonance Imaging of Metabolism

*a) Hyperpolarized carbon 13 ($^{13}C$) MRI:* Hyperpolarized carbon 13 ($^{13}C$) MRI allows the real-time *in vivo* non-invasive imaging of specific metabolic pathways. This non-radioactive technique detects hyperpolarized carbon compounds created through a dynamic nuclear polarization method that enhances their signal >10,000 fold.[301] Since carbon is the backbone of most organic molecules, different biomolecules labelled with $^{13}C$ have been used as exogenous CAs to dynamically detect their distribution throughout the body and to interrogate the activity of metabolic and physiologic processes. Hyperpolarized $^{13}C$ MRI can be used either for MR spectroscopy or for dynamic imaging which allows the generation of metabolic maps. One of the most widely used hyperpolarized compounds is $^{13}C$-pyruvate which is utilized to study glycolysis, mitochondrial energy production and amino acid synthesis. After parenteral injection, $^{13}C$-pyruvate is quickly taken up by cells, where it will be either: converted into lactate in the cytosol of tumour cells employing the Warburg effect or, oxidized and transported into the mitochondria to enter the TCA cycle to ultimately generate ATP or, channelled towards amino acid synthesis.[302] $^{13}C$-pyruvate has primarily been used in preclinical studies; however, its clinical application is increasing as reported by *Larson et al.*, who successfully performed dynamic hyperpolarized $^{13}C$-pyruvate MRI to quantify pyruvate-to-lactate metabolic conversion in prostate cancer patients.[303] Another exciting clinical application was an *in vivo* evaluation of non-small cell lung carcinoma (NSCLC) metabolism using intra-operative infusions of $^{13}C$-glucose to compare the metabolism between tumours and benign lung. Interestingly, alterations in glycolysis, glucose oxidation and consumption/oxidation of lactate showed heterogeneity within and between tumours, that correlated with the level of perfusion in the TME .[304] In a preclinical xenograft mouse model of renal cell carcinoma, *Dong et al.* used $^{13}C$-pyruvate MRI to predict which tumours would respond to everolimus and temsirolimus, two mTOR inhibitors, since the PI3K/mTOR axis is known to control glycolysis. Importantly, metabolic imaging can show changes in glycolytic flux within 24 h of treatment.[305] Other $^{13}C$ probes used to assess metabolic or physiologic processes include $^{13}C$-bicarbonate to interrogate pH,[306] $^{13}C$-acetoacetate as an indicator of mitochondrial redox status, $^{13}C$-urea to measure perfusion[307,308] and, $^{13}C$-fumarate for cellular necrosis.[302,309] Interestingly, several $^{13}C$ probes can be combined to perform multiple hyperpolarized MRI measurements. For instance, *Chen et al.* applied a three-dimensional dynamic dual-agent approach with $^{13}C$-pyruvate and $^{13}C$-urea to study differences in perfusion and metabolism between low- and high-grade tumours in a transgenic mouse model of prostate cancer.[309,310] In addition, the simultaneous use of different $^{13}C$ probes can allow ratiometric measurements[309,311] to determine response to treatment as *Datta et al.* showed in a rat model of glioma, where they used the ratio of lactate:bicarbonate to predict the response of these tumours to an anti-vascular endothelial growth factor antibody.

*b) Spectroscopy and TME:* Magnetic resonance spectroscopy (MRS) is a specialized non-invasive analytical technique that can generate a metabolic profile of normal tissues and their diseased counterparts to assess multiple conditions such tumours, ischemic stroke, and neurodegenerative diseases.[312] *In vivo* MR-based $^1H$ localized spectroscopy uses Point RESolved Spectroscopy Sequence (PRESS) or STimulated Echo Acquisition Mode (STEAM) programs to measure the signal from protons in an anatomical region of interest. A spectrum is generated in which the frequency shift of each molecule, relative to that of water, determines the identity of specific metabolites and the area under the peak specifies their concentration. MRS can be performed on a single voxel (a volume in three-dimensional space), termed single voxel spectroscopy, or the technique can use two/three phase-encoding directions to create a multi-voxel 2/3-dimensional spectra to produce maps of tissue chemistry, a technique that falls under the broad category of chemical shift imaging. The more popular Magnetic Resonance Spectroscopic Imaging (MRSI) sequences today are chemical shift imaging (CSI) [313] and Proton Echo-Planar Spectroscopic Imaging (PEPSI)[314] in the human brain. These techniques do not typically render high resolution spectra but are able to rapidly collect spectra over several localized areas to generate two-dimensional and three-dimensional chemical maps throughout a large tissue sample.

$^1H$ spectroscopy is sensitive to several products of enzymatic activity including: acetate, acetone, alanine, aspartate, betaine, choline, citrate, creatine, glucose, carnitine, phosphorylethanolamine, $\gamma$-amino-n-butyric acid, glutathione, glycine, glycerophosphorylcholine, glutamate, glutamine, lactate, myo-inositol, N-acetyl-asp-glutamate, N-acetyl-L-aspartate, phosphocreatine, phosphorylcholine, pyruvate, scyllo-inositol, succinate and taurine.

The $^1$H spectral peaks containing citrate, choline and lactate have been correlated with anaerobic metabolism as a marker of malignancy. An elevated choline spectrum, which corresponds to phosphocholine and other choline-containing compounds, termed total choline (tCho)[315], denotes an elevated choline metabolism resulting from enzymes that are deregulated in cancer. These can also affect glycolytic pathways including glucose transport, glycolytic flux, and lactate production. Thus, lactate and choline are useful biomarkers of malignancy, of aggressiveness and therefore, of tumour staging and of early response to therapeutic interventions.[316] Years ago, we interrogated the usefulness of MRS in the evaluation of prostate cancer in a transgenic mouse model (heterozygous *Pten* knockout combined with constitutively active ErbB-2 overexpression) and found that both choline and citrate were adequate biomarkers to distinguish prostate tumours from normal controls **(Figure 11)**.[317] mouse (bottom); data was processed using 8 Hz line broadening. (Cho:choline, Cit:citrate, mI:myo-inositol, Lac:lactate, Lip:lipid, Glx: glutamine and glutamate).[317]

Our findings have been supported in clinical studies of prostate cancer; an example of this is a report in which a quantitative MRS analysis of metabolite ratios of citrate, choline, creatine and spermine was performed.[318,319] We also employed this approach in a spontaneous mouse model of medulloblastoma, (MB), a predominantly paediatric brain tumour, that overexpresses constitutively active Smoothened receptor. We first assessed MB tumour formation by MRI and MRS to define the anatomical and metabolic alterations in brain tissue as a result of tumour progression.[320]

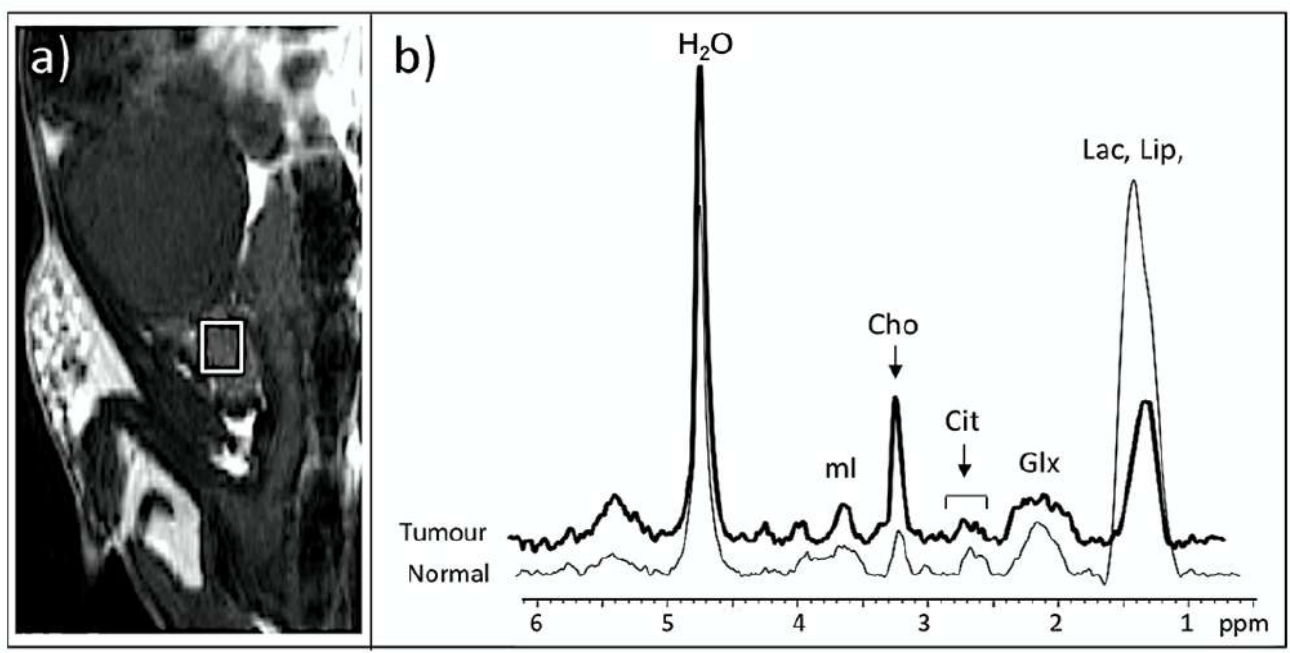

**Figure 11 – Single voxel $^1$H MRS.** Spectroscopy was performed using PRESS with TE: 11.7 ms, TR: 2000 msec, 4,000 transients, spectral width: 3 kHz and 512 k complex data points. **(a)** Example of actual placement of a voxel on the prostate (~~8 ml~~ total volume = 8 μL). **(b)** Representative spectra from a PB-ErbB-2D + Pten+/- (transgenic) mouse ~~(top) and a normal.~~ with a prostate tumour (thick line) and a control mouse with a normal prostate (thin line).[317]

In addition, following anatomical MRI to localize MB lesions, MRS established that treatment with a novel CDK1/CDK2 inhibitor, VMY-1-103, resulted in an increase in taurine and myo-inositol in the treated tumours. Importantly, this pattern correlated with molecular findings such as cell death and cyclin D1 down-regulation **(Figure 12)**.[321]

*In vivo* MRS measurements can be correlated with the *ex vivo* spectroscopy of tissues using high-resolution (HR) magic angle spinning (MAS),[318] which can be both accurate and reliable if strict sample preparation procedures are followed, as shown by *Opstad* with adult gliomas and by *Mazuel* who quantified the lactate concentration accurately with *ex vivo* MRS.[318,322]

One of the weaknesses of MRS is its inability to distinguish the biochemistry of the tumour versus that of the surrounding extracellular component, reason why many investigations use multi-parametric imaging approaches in order to bring more clarity in the interpretation of results to explain underlying processes.

Lastly, spectroscopic acquisitions can also be performed for non-proton-based nuclei, such as [19]F, [13]C and [31]P; the design of diverse biomolecules for this purpose, particularly in the field of hyperpolarized [13]C MRS, has allowed the study of specific metabolic pathways with a much higher sensitivity and specificity, resulting in the exponential growth of applications of this method in the study of human disease.

### c) CEST MRI in tumour metabolism

In tumours, glucose metabolism is altered at many levels, including an increased cellular glucose uptake by overexpressed glucose transporters GLUT1 and GLUT2. Additionally, the predominance of anaerobic glycolysis in cancer is caused by deregulation of glycolytic enzymatic activity including upregulation 6-phosphofructokinase, pyruvate kinase, hexokinase 2, glyceraldehyde-3-phosphate dehydrogenase, triose-phosphate isomerase, phosphoglycerate kinase 1 and enolase 1. Lactate dehydrogenase A (LDHA), which converts pyruvate into lactate, also causes a marked accumulation of lactate.

As mentioned previously, glucose and lactate can be determined via [13]C hyperpolarized MRI, MR spectroscopy and CEST-MRI. CEST-MRI mapping of glucose (glucoCEST) is possible due to the fast exchange rate of protons between the five hydroxylic groups of this sugar with water at a resonance of 1–1.2 ppm at high magnetic fields.[318,319] GlucoCEST is quickly gaining prominence due to its ease of use and the lack of ionizing radiation (in contrast to FDG-PET). *In-vivo* glucoCEST involves either the imaging of endogenous glucose in areas of high glucose consumption such as brain or the use of injected glucose to enable dynamic imaging. One disadvantage of using exogenous glucose as a CA, is its rapid metabolism, which limits imaging time. As with PET imaging, non-metabolizable glucose such as 2-Ddeoxy-D-glucose (2DG), exists in cells longer but presents a higher toxicity. A molecule that shows promise is 3-O-methyl-D-glucose (3OMG), which is considered nontoxic and has been used in the successful imaging of mammary gland tumours in a mouse model of breast cancer .[276,323]

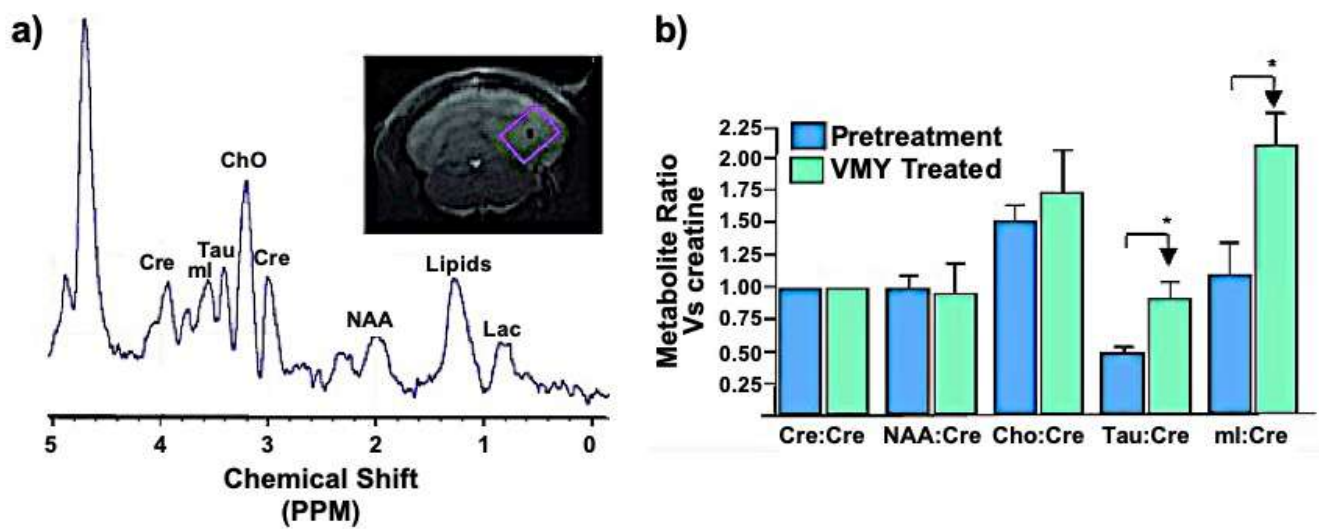

**Figure 12 – Effect of VMY on tumour growth and metabolism. (a)** Representative voxel placement and *in vivo* MR spectral profile of medulloblastoma; **(b)** Effect of VMY on metabolite ratios. Metabolite ratios were established relative to creatine (Cre). (Abbreviations: Cre:creatine, Tau:taurine,NAA:N-acetyl-aspartate). Data are average ± standard deviation of n=3 mice (asterisk), p<0.05 using one-tailed Mann Whitney U-tests.[320,321] MRS was performed using PRESS with TE: 20 ms, TR: 2500 ms, averages: 1,024, spectral width of 4 kHz, 2,048 complex data points and 6 Hz line broadening, using a voxel of 1–2 mm on edge located entirely in tumor areas avoiding contamination from normal brain tissue. [320,321]

Increased glucose detection co-registers with tumorous areas due to their increased uptake as a result of enhanced metabolic activity but can also indicate a higher vascular permeability or increased perfusion, both of which are characteristic of a tumour phenotype. We performed multiparametric (anatomical, MRS, DWI and glucoCEST) MR imaging on a transgenic mouse model of medulloblastoma **(Figure 13)**, where we detected a highly altered glucose metabolism in areas of tumour compared to normal brain tissue **(Figure 13d)**.

### 4.4 MRI to visualize vascularization in TME

### a) Arterial spin labelling (ASL) MRI: Arterial spin labelling (ASL) is a functional MRI technique that visualizes the flow of blood in the vasculature, with the added advantage that it

does not require the administration of exogenous CAs. In ASL, arterial blood is excited with a spatially selective radio frequency pulse that is outside the imaging field of view. As the labelled blood flows towards the tissue through capillaries and into the extravascular space, the images are acquired with a rapid imaging technique.[324] By imaging a region of interest both before and after ASL, the differential image intensities acquired enable the quantification of flow rates over time.[325] The two main parameters measured by ASL are tissue blood flow and arterial transit time, which are highly reproducible and repeatable but exhibit low spatial resolution. Since ASL is sensitive to motion, it is particularly suited for imaging brain where this problem is minimized. For this reason, ASL has been used to: assess cerebral tumour diagnosis and grade, to aid in the differentiation of gliomas from lymphomas[326] to monitor of response to anti-angiogenic agents and to monitor of responses to radiation treatment in the clinic, with a notable emphasis on pediatric tumors.[327] It is also clear that ASL is highly applicable to the study of the TME in mouse models.[77] Lastly, ASL has been shown to analyse tumour perfusion in extra-cerebral sites such as head and neck cancers,[328] and pancreatic and renal tumours.[329]

*b) Diffusion MRI:* The random diffusion of water molecules in biological organisms is restricted by tissue microstructures such as cellular membranes, fibrosis and macromolecules and is therefore, a measure of cellularity and cellular membrane integrity. The impedance of movement of water protons within cell wall boundaries can be quantified through the apparent diffusion coefficient (ADC), allowing the generation of spatial maps of diffusion, among several other parameters. Thus, water diffusion and ADC are known to correlate with vascular microstructure, BBB integrity (when imaging brain) and tumour cellular density, making them sensitive biomarkers of tumour malignancy and prognosis. The blood in the tumour microvasculature is a prominent compartment that is reflected directly in diffusion weighted MRI (DWI) measurements; thus, DWI also interrogates tissue perfusion. Given that the cellular composition in tumorous tissue can be vastly different from normal tissue, especially in the case of highly desmoplastic tumours such as pancreatic ductal adenocarcinoma, diffusion anisotropy measurements are essentially a measure of tissue cellular organization and therefore, a measurement of the TME. *Granata et al.* for instance, showed the capability of DWI to differentiate normal pancreas from peri-tumoural inflammation and pancreatic tumour.[330–332] In a mouse model of the brain cancer medulloblastoma, we showed a highly abnormal DWI and ADC values between the tumour and normal tissue **(Figure 13c)**. DWI can detect changes resulting from tumour response to therapies weeks before changes in tumour volume are detected.[333]

Tissues that exhibit a highly organized microstructure restrict water movement in certain directions. Examples of this are white matter tracts in the central nervous system, glandular ducts in the mammary gland or collagen fibres in tumours. Fractional anisotropy (FA), a measure of this directionally constrained water diffusion, can be evaluated by diffusion tensor imaging (DTI), an extension of DWI. In DTI, the movement of water is assessed in six or more directions. Thus, tumours characterized by a high degree of cellularity, fibrosis or invasion of neuronal axons, will alter the inherent tissue FA, enabling the discrimination of normal versus tumorous tissue and benign versus malignant masses. *Kakkad et al.* showed that ADC and FA correlated with tumour hypoxia and necrosis in a mouse model of breast cancer[263] and *Luo et al.* combined DTI, DWI and DCE to increase the specificity of MRI to diagnose benign versus malignant breast tumours with a high specificity.[334]

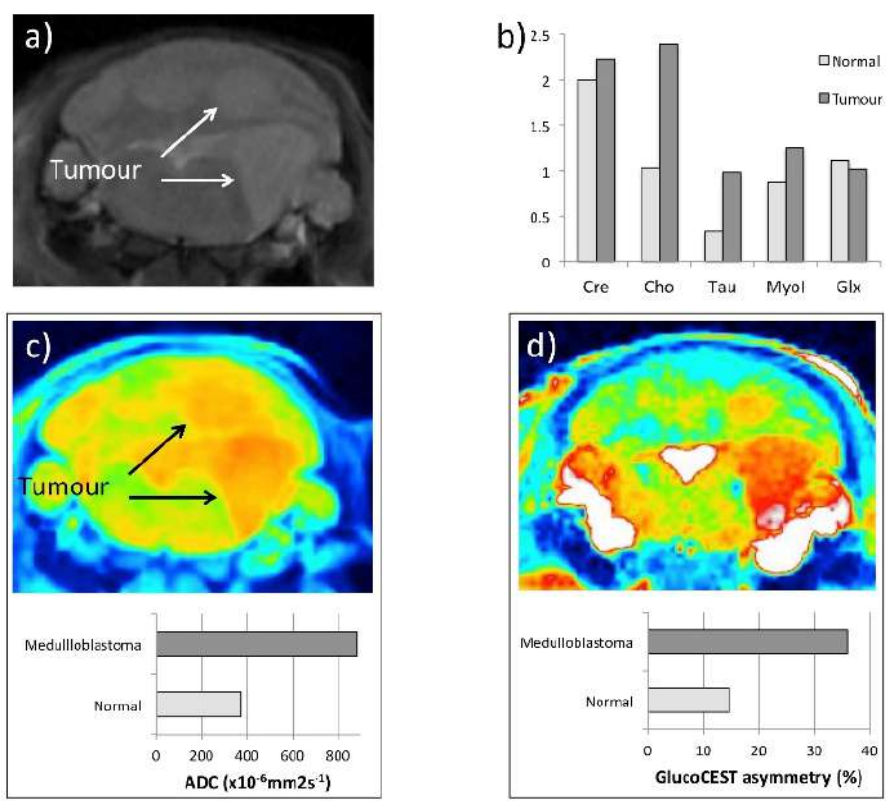

**Figure 13 – MRI of mouse medulloblastoma (MB).** **(a)** Anatomical T2-weighted MRI of mouse brain tumour (RARE sequence with TR: 3000 ms, TE: 12 ms, matrix: 256, and RARE factor); **(b)** Comparison of $^1$H MRS spectra of mouse MB versus healthy brain (PRESS with TR: 2000 ms, TE: 20, averages: 1024); **(c)** Diffusion-weighted imaging of mouse MB and differential ADC values between tumour and healthy brain (SPIN ECHO sequence with TR: 2300 ms, TE: 27, matrix: 128, averages: 2 and four different B values); **(d)** GlucoCEST spatial map (MTRasym at ~1.1 ppm) and comparison of the mean ROI values of GlucoCEST$_{asym}$ between MB and healthy brain (RARE pulse sequence with TR: 24, TE: 8, averages: 2, repetitions: 32, matrix 122x128).

In the assessment of perfusion, the combination of DWI and DCE-MRI can significantly increase the performance of MRI in the assessment of perfusion. This was demonstrated by *Koh et al.* who evaluated patients with metastatic brain tumours in order to interrogate their recurrence; importantly, this method was capable of distinguishing tumour recurrence from the effects of therapy with high specificity.[335] This increased diagnostic accuracy was also proven by *Zheng et al.* to discriminate parotid tumours.[336]

### c) Dynamic-susceptibility contrast (DSC) and dynamic-contrast enhanced (DCE) MRI:
Dynamic-contrast enhanced (DCE) and dynamic-susceptibility contrast (DSC) MRI are imaging techniques that can be used to interrogate inadequate blood perfusion, a cause of hypoxia.[337] By defining the differences between tumour diffusion- and perfusion-weighted MR images, the "salvageable tissue" areas within and around tumours can be defined. Perfusion images are used to determine the blood saturation of a tissue. Perfusion is also a measure of cellular organization. Since the imaging output from both diffusion and perfusion are manifestations of the physical properties of tissue, the combination of both imaging techniques enhances the sensitivity and specificity in the evaluation of tumours in response to therapeutic intervention.[338]

### DCE- MRI
Clinical DCE-MRI (mentioned above in the study of hypoxia) involves primarily the use of low molecular weight (LMW) T1 CAs such as gadopentetate dimeglumine or gadobutrol, both used extensively in humans. These agents extravasate rapidly into the extracellular space of both normal and tumorous tissue and can therefore, present challenges in the high-resolution detection of the enhanced permeability present in many tumour models. To address these challenges, a wide variety of macro-molecular weight agents (MMW) are available for preclinical research. The MMW CAs remain in the bloodstream longer and predominantly leak across the hyper-permeable and aberrant tumour vasculature, allowing changes between perfusion, permeability and blood volume to be better defined.[339,340] MMW agents for DCE-MRI are typically gadolinium-based high polymer CAs and small molecular CAs that bind serum albumin. However, since some MMW CAs exhibit toxicities due to accumulation in tissues and their capacity to trigger abnormal immune responses, this precludes them from being used in the clinic.

Nanoparticles containing either T1 CAs such as Gadolinium or manganese, or T2 agents such as $Fe_3O_4$, show promise for angiogenesis studies with the added advantage that they can be multi-functionalized through the incorporation of fluorophores, drugs, etc. However, they tend to be taken up by the mononuclear phagocytes, can be retained in tissues and are less effective as blood pool agents versus LMW CAs.[341]

The differential tumour permeability exhibited by MMW CAs allows for discriminating between low- and high- grade tumours, which is generally not possible with the LMW CAs.[342–344] In addition, numerous studies have looked into the reliability of DCE-MRI to evaluate the response of tumours to anti-angiogenic therapy, with MMW agents bigger than 17 kDa capable of detecting a precipitous decline in $K^{trans}$ (a measure of capillary permeability) and of the fractional plasma volumes as soon as 24 h after treatment.[345] These characteristics of MMW CAs make them particularly useful for longitudinal preclinical studies of the TME.

*DSC-MRI*

Dynamic susceptibility contrast (DSC) MRI is a minimally invasive imaging technique that involves the use of T2*-weighted imaging sequences (for instance, gradient-echo echo-planar sequences) to measure the signal loss after paramagnetic, non-diffusible gadolinium-based CAs travel through tissues. After an intravenous bolus injection, the paramagnetic CAs flow through the capillaries, and fast imaging sequences are rapidly and repetitively performed yielding quantitative spatial maps of: blood flow, blood volume, mean transit time (MTT) and time to peak (TTP) and, whose values are proportional to the CA concentration.[346]

DSC-MRI is therefore suited for imaging of brain tumours. In the healthy brain, the CA is restricted to the vascular space by the BBB. However, the BBB is compromised by brain tumours such as medulloblastoma and glioblastoma multiforme**,** enabling the leakage of the DSC CAs into the extracellular space, which is easily discernible by T1-MRI. DSC-MRI (particularly in combination with DCE-MRI) allows to determine tumour grade, predict tumour response to treatment and distinguishes tumour recurrence from necrosis as a result of radiotherapy.[347,348]

*4.5 Multiparametric MRI of the TME*

The TME exhibits an inter- and intra-tumoural heterogeneity that is highly spatially and temporally dynamic. This not only makes an accurate diagnosis difficult, but also hinders the prediction of how tumours will respond to treatment, complicating the assessment of prognosis and affecting the overall outcome. The MRI techniques discussed so far are useful for non- or minimally-invasively assessing the TME; however, while each technique can help characterize important aspects of tissue physiology, in isolation they may provide an incomplete assessment of the tumour, and may exhibit false-positive or -negative results. To minimize this, multiparametric MRI (mpMRI), that is, the combination of multiple MR imaging techniques to measure several TME parameters, is increasingly used. mpMRI generates complementary information of tumour physiology, increasing the specificity and sensitivity of MRI, improving the accuracy of imaging diagnostics and enabling a more precise characterization of the TME.

Several research groups have designed a variety of MRI-based molecular imaging paradigms, paradigms that are constantly improving, aided by increased magnetic field strengths, advancements in imaging protocols and importantly, in imaging analytical methods, that include algorithms that can enable computer learning and thus, permit more accurate and objective evaluations. Exciting examples of mpMRI to study the TME include: a) the integrated analysis of TME microstructure (DWI) and perfusion (DSC) to cluster glioblastoma patients, a stratification that is correlated with invasiveness and survival;[349] b) the combination of DWI and DTI to distinguish benign from malignant breast cancer lesions;[334] c) the assessment of the tumour versus stromal ratios to characterize the invasiveness of breast cancer tumours[350] by using short-tau inversion-recovery (STIR) $T_2$ weighted imaging, diffusion-weighted imaging (DWI) and post-contrast dynamic imaging; d) the use of $^{13}CUrea$ and $^{13}Cpyruvate$ for perfusion- and MRSI-$^{13}C$ hyperpolarized MRI, respectively, in order to study lactate metabolism and its correlation with tumour progression in the TRAMP mouse model of prostate cancer;[351] e) the assessment of dynamic changes of pancreatic cancer progression in the KPC mouse model through T1- and T2-weighted MRI, CEST-MRI, and DWI;[352] f) the hypoxic and vascular niches in the TME of human prostate tumours, as determined by the combination of DWI, BOLD, and DCE were correlated to Gleason score[264] and lastly; e) an ambitious user-independent TME mapping approach fused imaging parameters of energy production and neovascularization to dissect the tumour heterogeneity of human glioblastoma, its subtypes and progression-free survival.[353]

# 5. Conclusions and outlook

Optical imaging and MRI are based on fundamentally different physicochemical principles, which accounts for their mostly non-overlapping strengths and shortfalls (mentioned above). This is why in the study of the TME, the information these technologies yield is mutually informative and complementary. Evidence of this abounds in various studies that have used these modalities in concert in the interrogation of the TME.[354,355] Interestingly, the differences between optical imaging and MRI have been exploited to design "smart" multi-functional nanoparticles that encapsulate targeting, diagnostic and therapeutic capabilities (theranostics), which incorporate optical imaging probes and paramagnetic particles among several other components, such as targeting moieties and drugs.[356–359]

Tumour microenvironment imaging modalities based on fluorescence and MRI have been evolving continuously. Fluorescence microscopy based TME imaging has several advantages due to its simplicity and tailorability. Acidic tumour microenvironment and hypoxia have been extensively studied and assessed using fluorescence microscopy. However, poor penetration depth of the photons due to scattering and absorption by tissue is one of the major limitations of fluorescence based TME imaging. Multiphoton imaging and use of longer wavelengths have solved the problem to some extent and more advances are still developing in this direction. Fluorescent based pH sensors sense either direct changes in a pH sensitive fluorophore or changes in energy transfer between a donor and acceptor. Additionally, research has also been shifting towards creating fluorescent based nanosensors that can offer more sensitivity and tailorable design. Thus, flexibility in design allows for enhanced specificity by decorating the nanosensors with affinity biomolecules, that in turn sequester specifically in region of interest in the body or organ. Although nanoprobes, proteins and gene-based sensors have shown promising results and are considered non-invasive, they still needs to be injected into the body, and proper delivery remains a challenge to be solved before fully realizing their traslational potential. The different properties that need improvement are circulation times, non specific localization and the assesement of long term cytotoxicity.

Historically, while MRI has been and remains a standard of care in the management of cancer and other clinical indications, the use of optical imaging in the clinic has been limited by the lack of FDA-approved optical imaging devices and target compounds, as well as its relatively shallow depth of penetration in tissues. This is rapidly changing commensurate with the development of exciting novel applications in which optical imaging has shown promise in both preclinical and clinical settings, such as fluorescence-guided intra-operative imaging and surgery where tumour margins can be established and residual tumours can be identified, based on the detection of TME biomarkers.[360–363]

Tumour hypoxia has been known to affect cancer metastasis and to perturb the effectiveness of drugs. Therefore, various new and more sensitive fluorescent dye molecules have been developed that employ fluorescence lifetime change as the indicator of dissolved oxygen. The basic structure of these fluorophores consists of Ruthenium–Chromium and Platinum phosphor porphyrins complexes. Similar to other fluorescent nanoprobes, oxygen-sensitive nanoprobes based on dyes have exhibited very high sensitivity, stability and selectivity. Therefore, it can be stated that fluorescent nanosensors hold the key for further growth in the field of fluorescence based TME bioimaging in years to come.

The capability of MRI to faithfully assess the pathophysiology of the TME can also be enhanced by combining it with other imaging modalities such as ultrasound, PET, CT, optical imaging, etc., an approach that is referred to as multimodal imaging, a topic beyond the scope of this review. Finally, most MRI techniques presented in this review can be enhanced using a multitude of exogenous compounds with a wide range of sizes and functionalities, generally known as nanotheranostics. This ever-growing field entails the use of nanoagents, designed for highly selective targeted delivery and improved biodistribution, in order to deliver therapeutic payloads (including chemotherapeutics, miRNAs, heat-sensitive probes, etc.) and for increasingly accurate diagnosis by incorporating a variety of magnetic and fluorescent CAs for multimodal imaging.

## Acknowledgements

We are grateful to the European Research Council (ERC) under the European Union's Horizon 2020 research and innovation programme (project No. 759959, INTERCELLMED), My First AIRC Grant (MFAG-2019, No. 22902), the FISR/MIUR-C.N.R., Tecnopolo di Nanotecnologia e Fotonica per la medicina di precisione (project No. B83B17000010001) and Tecnopolo per la medicina di precisione - Regione Puglia (project No. B84I18000540002). The work herein was further supported by NIH-P30 CA051008-25, NIH-S10 OD 025153-01A1, NIH R44AG066386-01 and DE-AC52-07NA27344. The MRI data presented was entirely performed in the Georgetown-Lombardi Preclinical Imaging Research Laboratory.